\documentclass[10pt,twocolumn,letterpaper]{article}

\usepackage{cvpr}              % To produce the CAMERA-READY version
\usepackage[accsupp]{axessibility}  % Improves PDF readability for those with disabilities.

\usepackage{graphicx}
\usepackage{amsmath}
\usepackage{amssymb}
\usepackage{booktabs}
\usepackage[dvipsnames]{xcolor}
\usepackage{multirow}
\usepackage{amsmath,bm}

\newcommand{\coefcb}{{c_b}}
\newcommand{\coefcr}{{c_r}}

\usepackage[pagebackref,breaklinks,colorlinks]{hyperref}

\usepackage[capitalize]{cleveref}
\crefname{section}{Sec.}{Secs.}
\Crefname{section}{Section}{Sections}
\Crefname{table}{Table}{Tables}
\crefname{table}{Tab.}{Tabs.}

\def\cvprPaperID{5551} % *** Enter the CVPR Paper ID here
\def\confName{CVPR}
\def\confYear{2022}

\begin{document}

\title{Practical Learned Lossless JPEG Recompression with Multi-Level Cross-Channel Entropy Model in the DCT Domain}

\newcommand{\authorinstitude}[1]{{\textsuperscript{#1}}}

\author{Lina Guo\authorinstitude{1}\authorinstitude{2}\thanks{Equal contribution.}\ ,
Xinjie Shi\authorinstitude{1}\footnotemark[1]\ ,
Dailan He\authorinstitude{1}\footnotemark[1]\ ,
Yuanyuan Wang\authorinstitude{1},
Rui Ma\authorinstitude{1},
Hongwei Qin\authorinstitude{1}, 
Yan Wang\authorinstitude{1}\authorinstitude{3}\thanks{Corresponding author. This work is done when Lina Guo, Xinjie Shi, and Rui Ma are interns at SenseTime Research.}\\
SenseTime Research\authorinstitude{1}, National University of Defense Technology\authorinstitude{2}, Tsinghua University\authorinstitude{3} \\
{\tt\small \{guolina1, shixinjie, hedailan, wangyuanyuan, marui, qinhongwei, wangyan1\}@sensetime.com} \\
{\tt\small guolina19@nudt.edu.cn, wangyan@air.tsinghua.edu.cn}
}
\maketitle

%%%%%%%%% ABSTRACT
\begin{abstract}
JPEG is a popular image compression method widely used by individuals, data center, cloud storage and network filesystems. However, most recent progress on image compression mainly focuses on uncompressed images while ignoring trillions of already-existing JPEG images. To compress these JPEG images adequately and restore them back to JPEG format losslessly when needed, we propose a deep learning based JPEG recompression method that operates on DCT domain and propose a Multi-Level Cross-Channel Entropy Model to compress the most informative Y component. Experiments show that our method achieves state-of-the-art performance compared with traditional JPEG recompression methods including Lepton, JPEG XL and CMIX. To the best of our knowledge, this is the first learned compression method that losslessly transcodes JPEG images to more storage-saving bitstreams.
\end{abstract}

%%%%%%%%% BODY TEXT
\section{Introduction}
\label{sec:intro}

JPEG \cite{wallace1992jpeg}, a popular image compression algorithm, is used by billions of people daily and JPEG images spread widely in data center, cloud storage and network filesystems. According to a survey, in operating network filesystems like Dropbox, JPEG images make up roughly 35\% of bytes stored \cite{horn2017design}. However, most of these images are not sufficiently compressed due to the limitation of JPEG algorithm: relying on hand-crafted module design and hard to eliminate data redundancy adequately. Actually, JPEG algorithm has been developed for many years so that it has already been outperformed by other more recent image compression methods, such as JPEG2000 \cite{rabbani2002jpeg2000}, BPG \cite{bellard2015bpg}, intra coding of VVC/H.266 \cite{ohm2018versatile} and deep-learning based methods \cite{balle2016end, balle2018variational, minnen2018joint}. However, these subsequent image compression methods devote to process original images in lossless format like PNG while ignoring the need of further compressing trillions of existing JPEG images losslessly.

Considering the recompression needs of storage service, there exist several methods on further compression of JPEG images, \eg Lepton \cite{horn2017design}, JPEG XL \cite{alakuijala2019jpeg, alakuijala2020benchmarking}, and CMIX \cite{cmix}. However, they rely on hand-crafted features and independently optimized modules, limiting compression efficiency. Along with the quick proliferation of mobile devices saving and uploading JPEG images, these storage systems have become gargantuan and existing JPEG recompression algorithms are not expected to be optimal and general solutions to storage challenges faced by service providers. 

%\begin{figure}[t]
  %\centering
  %\includegraphics[width = .4\textwidth]{img/overview.pdf}
  %\caption{Overall architecture of the proposed JPEG lossless recompression method}
  %\label{fig:frame}
%\end{figure}

We propose an efficient JPEG image lossless recompression neural network using quantized DCT \cite{ahmed1974discrete} coefficients as input, which is stored in the JPEG file. To the best of our knowledge, this is the first work proposing a deep learning based model dedicated for lossless recompression of JPEG images %, achieving losslessly compresses an average JPEG images by at least 30\% 
and outperforms existing traditional methods including Lepton, JPEG XL, and CMIX by a large margin.

\begin{figure*}[t]
  \centering
  \includegraphics[height=4.5cm]{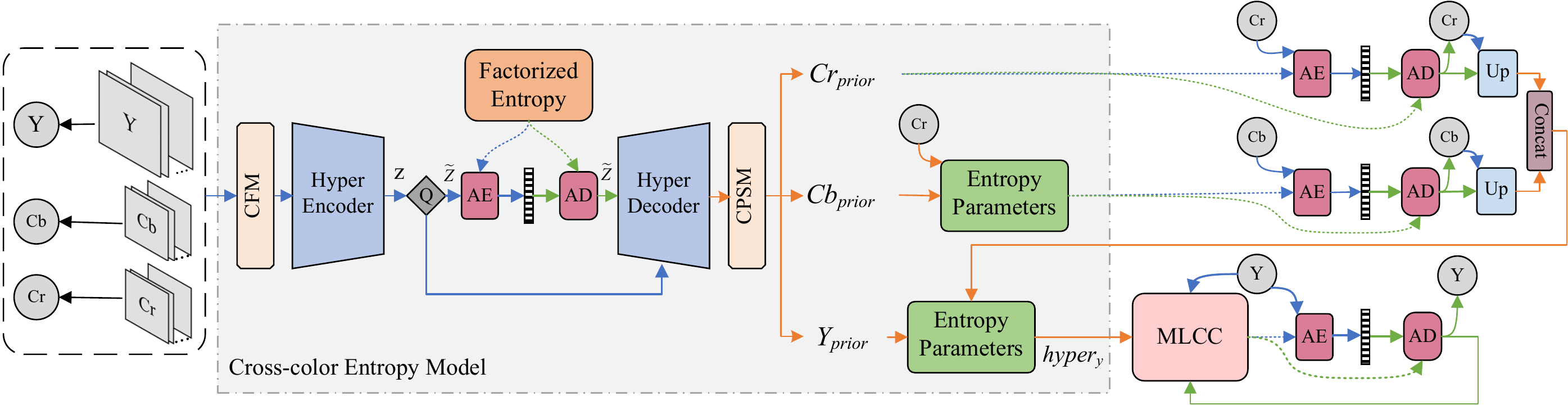}
  \caption{Overall architecture of the proposed JPEG lossless recompression method. AE and AD stands for arithmetic coding and arithmetic decoding respectively. Blue and green lines indicate data-flow for encoding and decoding respectively, orange lines are shared.}
  \label{fig:frame}
\end{figure*}

% todo
In our method, JPEG in YCbCr 4:2:0 format is considered because of its popularity. %, where Cb and Cr components are subsampled but Y is retained at the same resolution and contains more information by considering the human visual system. 
As shown in \cref{fig:frame}, we first construct a color-space entropy model for YCbCr 4:2:0 format which extracts side information $ z $  as prior to build conditional distribution of the three components. Then we further exploit the correlation of Y, Cb, and Cr components sequentially (\ie Cb component conditioned on Cr, Y component conditioned on both Cb and Cr). Additionally, since Y component is much more informative than Cb and Cr components, we propose a Multi-Level Cross-Channel (MLCC) entropy enhancement model for Y component to reduce the mismatch between estimated and true data distribution. %All modules of our network are end-to-end optimized by gradient decent.  %which has long been proven to be easier to learn more informative feature and estimate more accurate probabilistic model resulting in better compression performance. 
%Our approach achieves state-of-the-art performance and reduces storage by more than 10\% compared with Lepton, the best traditional JPEG lossless recompression algorithm. %todo

In conclusion, our main contributions include:
\begin{itemize}
    \item We propose an end-to-end lossless compression model for images already encoded with JPEG format. To the best of our knowledge, this is the first approach investigating learning-based JPEG recompression, which further benefits the widely adopted JPEG format based on powerful  data-driven techniques.
    \item Experiments show that our proposed JPEG recompression method achieves state-of-the-art performance, outperforming Lepton, JPEG XL and CMIX. Also, our model has reasonable running speed and is a promising candidate for practical JPEG recompression. %Our method also demonstrates excellent generalizability on all quality levels with acceptable computational efficiency.
\end{itemize}

%%%%%%%%%%%%%framework picture%%%%%%%%%%%%%%%%%

\section{Related work}
\label{sec:relate}
\subsection{Overview of JPEG algorithm}
\label{ssec:jpeg}
JPEG algorithm first converts image from RGB sources to YCbCr color space (one luma component (Y) and two chroma components (Cb and Cr)). Then, considering human visual system is far more sensitive to brightness details stored in the luma component than to color details stored in two chroma components, the luma component is supposed to be more important than chroma components. Most JPEG images adopt YCbCr 4:2:0 format, where Y retains the same resolution while Cb and Cr components are subsampled as $\frac{1}{4}$ of their original resolution. Next, every component is divided into $8\times8$ pixel blocks and each pixel block is transformed by discrete cosine transform (DCT) into a matrix of frequency coefficients (DCT coefficients) of the same size. Subsequently, these three components are quantized by two quantization tables: Y component uses a quantization table while Cb and Cr components share another quantization table. Finally, all DCT coefficients are compressed by lossless Huffman coding \cite{huffman1952method}. Importantly, in order to deploy Huffman coding, the two-dimensional DCT coefficients have to be turned into a one-dimensional array. Zig-zag scanning is adopted here to group similar frequencies together to obtain better performance. % of Huffman coding.

\subsection{JPEG recompression methods}
\label{ssec:recompression}
\textbf{Lepton} \cite{horn2017design} proposed by Horn \etal, which achieves 22\% storage reduction after recompressing JPEG losslessly, mainly focuses on optimization of entropy model and symbol representations. Instead of using Huffman coding, Lepton uses more efficient arithmetic coding \cite{10.1145/214762.214771}. Further more, combining unary, sign and absolute value, the representation method in Lepton outperforms other encodings like pure unary and two's complement with fixed length. Lepton also deals with DC component by predicting it from AC components and storing the residual.

\textbf{JPEG XL} \cite{alakuijala2019jpeg, alakuijala2020benchmarking} is a versatile compression method supporting both lossless and lossy compression. For existing JPEG images, losslessly transcoding them to JPEG XL is also supported. JPEG XL achieves better compression ratio by extending the $8\times8$ DCT to variable-size DCT which allows block size to be one of 8, 16 or 32. Besides, JPEG XL uses Asymmetric Numeral Systems \cite{duda2015use} in place of Huffman coding. Instead of using fixed quantization matrix globally, the quantization matrix in JEPG XL can be scaled locally to better accommodate the complexity in different areas. Compared with the primitive DC coefficient prediction mode in JPEG, JPEG XL supports eight modes and will choose the mode producing the least amount of error.

\textbf{CMIX} \cite{cmix} is a general lossless data compression program aimed at optimizing compression ratio at the cost of high CPU/memory usage. It achieves state-of-the-art results on several compression benchmarks. CMIX uses an ensemble of independent models to predict the probability of each bit in the input stream. The model predictions are combined into a single probability using a context mixing algorithm. The output of the context mixer is refined using an algorithm called secondary symbol estimation (SSE). CMIX can compress all data files losslessly, including JPEG images.

\subsection{End-to-end image compression}
\label{ssec:learned}
%Recent years, with the breakthrough achievements of deep neural networks in many vision tasks, lots of deep-learning based image compression method has subsequently emerged. Different from traditional method which rely on hand-craft feature and adopt separate optimization strategy at each module, deep-learning based method deploy neural network to learn feature automatically and jointly jointly optimize modules through end-to-end training, which has long been proven to be easier to learn more informative feature and eliminate data redundancy resulting in better compression performance. According to whether there is information loss, learned image compression approaches are classified two types: lossy and lossless.

\textbf{Learned lossy compression.} Since Ball\'e \etal \cite{balle2016end} proposes an end-to-end learned image compression method based on variational autoencoder \cite{vincent2008extracting} (VAE) architecture, subsequent deep-learning based approaches continue to explore and improve similar architectures (\eg \cite{ balle2018variational, rippel2017real, minnen2018joint, lee2018context, cheng2020learned, hu2020coarse,  guo2021causal, yuan2021learned, he2021checkerboard,chen2021end, yuan2021block}).  These methods initially focus on how to deal with non-differential quantization and rate estimation to enable end-to-end training \cite{balle2016end, rippel2017real}. Then, in order to build more accurate entropy model to further reduce the cross entropy (corresponding to bit rate), some methods \cite{balle2018variational, hu2020coarse} devote to introduce hyperprior models to the VAE architecture. More recent approaches investigate context models for more accurate entropy estimation, \eg adding pixel-wise \cite{minnen2018joint} or channel-wise autoregressive \cite{minnen2020channel} modules. These techniques have greatly improved the performance of learned image compression. Actually, all of the learned methods mentioned above have outperformed JPEG. The performance of newest methods \cite{guo2021causal,yuan2021block} even surpasses the intra coding of latest standard VVC/H.266 \cite{ohm2018versatile}. However, they focus on the compression of images stored in lossless format like PNG, serving as replacement of JPEG instead of recompressing existing JPEG images.
 
 \textbf{Learned lossless compression.} Our research is more related to learned lossless image compression. In theory, any probabilistic model can be used together with entropy coders to compress data into compact bitstreams losslessly. The bit rate lower bound is given by the probabilistic model according to Shannon's landmark paper \cite{Shannon1948A} (\ie the more accurate the probabilistic model is, the lower the bit rate will be). %Therefore, learned lossless compression can be achieved by using several popular probability estimation models, such as 
 Representative learned lossless image compression methods include likelihood-based generative models (\eg PixelCNN \cite{oord2016conditional}, PixelRNN \cite{van2016pixel}, MS-PixelCNN \cite{reed2017parallel}), bits-back methods (\eg BB-ANS \cite{townsend2018practical}, Bit-Swap \cite{kingma2019bit}, Hilloc \cite{townsend2019hilloc}) and flow-based models (\eg IDF \cite{hoogeboom2019integer}, IDF++ \cite{van2020idf++}, iVPF \cite{zhang2021ivpf}). To reduce computation complexity, a parallelizable hierarchical probabilistic model is proposed in L3C \cite{mentzer2019practical}, which is the first practical full-resolution learned lossless image compression method. This hierarchical probabilistic modeling idea is later improved by SReC \cite{cao2020lossless} and the multi-scale progressive statistical model \cite{zhang2020lossless}. Nevertheless, same as learned lossy image compression, these learned lossless compression methods still only consider images stored in PNG format while ignoring the vast already existing JPEG images. In our investigation, we find these methods cannot be used directly to losslessly compress JPEG images, which we focus on in this work.

%------------------------------------------------------------------------METHOD------------------------------------------------------------------
\section{Method}
\label{sec:meth}
%Our goal is design a deep learning model capable to compress JPEG, which reduce storage redundancy of JPEG in data center and cloud storage. YUV 4:2:0 is adopted as input-output format in JPEG and consists of subsampled chroma components(U and V), while the luma component(Y) retained at the same resolution without subsamlped. Usually, Y channel is more informative compared to U and V since human perception is far less sensitive to color details (captured by the chroma components) than to brightness details (in the luma component). 
%Our method design a efficient entropy model in the DCT domain capable to compress JPEG.  Since JPEG adopt YUV 4:2:0 format and Y channel contains more information than U and V.  Therefore, we follow a stage approach: first, a transform framework is adopted to align shape of DCT coefficients, and then aligned components is sent to a called hyperprior networks proposed in [] used to help the conditional of all components. Subsequently,  our model specifies conditional of components one-by-one. Importantly, multi-scale cross-channel entropy enhancement model is used to further accurately predict the distribution of Y channel.

\subsection{Framework}
\label{ssec:frame}

The overall framework of the proposed model is presented in \cref{fig:frame}. Since DCT domain is used in our method to design an efficient entropy model, we first rearrange each $8\times8$ block of DCT coefficients in order to learn better distribution (\cref{ssec:rearrangement}). Because JPEG usually adopts YCbCr 4:2:0 format, we apply a Coefficient Fusion Model (CFM) detailed in \cref{ssec:cem} to align the shape of DCT coefficients from different color components. After shape alignment, DCT coefficients are sent to a \emph{Hyper Encoder} and will produce the hyperprior $ \widetilde{z}$, which is saved in the bitstream as side information. 
Subsequently, the coding prior of the three color components will be obtained after the hyperprior $ \widetilde{z}$ goes through \emph{Hyper Decoder} and a Coefficient Prior Split Model (CPSM) detailed in \cref{ssec:cem}. 

Besides the shared hyperprior $ \widetilde{z}$, we further reduce the statistical redundancy by explicitly modelling the correlation between color components and DCT coefficients. Detailed in \cref{ssec:cem}, we estimate Cr distribution conditioned on  $ \widetilde{z}$, Cb distribution conditioned on both $ \widetilde{z}$ and Cr, and Y distribution conditioned on $ \widetilde{z}$, Cr and Cb. According to human perception, Y component contains more information than Cb and Cr components. We propose a multi-level cross-channel (MLCC) entropy enhancement model to better predict Y distribution, which is described in \cref{ssec:MLCC}.

Finally, we use arithmetic coding \cite{10.1145/214762.214771} based on these probabilistic distributions to compress component coefficients into compact bitstreams losslessly.

%\subsection{Entropy coding and probability model}
%\label{ssec:distribution}
%Lossless compression technique is essentially an entropy coder along with probability model. JPEG algorithm adopts Huffman coding together with Huffman tables defining the probability model to compress quantized DCT coefficients losslessly. However, these coefficients are considered to be independently and identically distributed (i.i.d.) under this fixed probability model, resulting in mismatch between estimated and true data distribution which decreases compression savings. 

%Our method contains two major improvements in this aspect. First, our method uses Laplace distribution with different parameters for each coding symbol to obatin an adaptive probability model, where the two parameters (scale $ b $ and location $ \mu $) of each Laplace distribution are learned by neural networks. Actually, it is known that AC coefficients of Fourier-related transforms, like DCT coefficients of JPEG algorithm, are from Laplace distribution \cite{Minguillo2001JPEG}. Second, we replace the Huffman coding with arithmetic coding, which allows compressing coding symbols into \cite{10.1145/214762.214771}.  %which allows varying distributions as a function of previously encoded symbols \cite{10.1145/214762.214771}.

\subsection{DCT Coefficients Rearrangement}
\label{ssec:rearrangement}

\begin{figure}[htb]
  \centering
  \includegraphics[width = .4\textwidth]{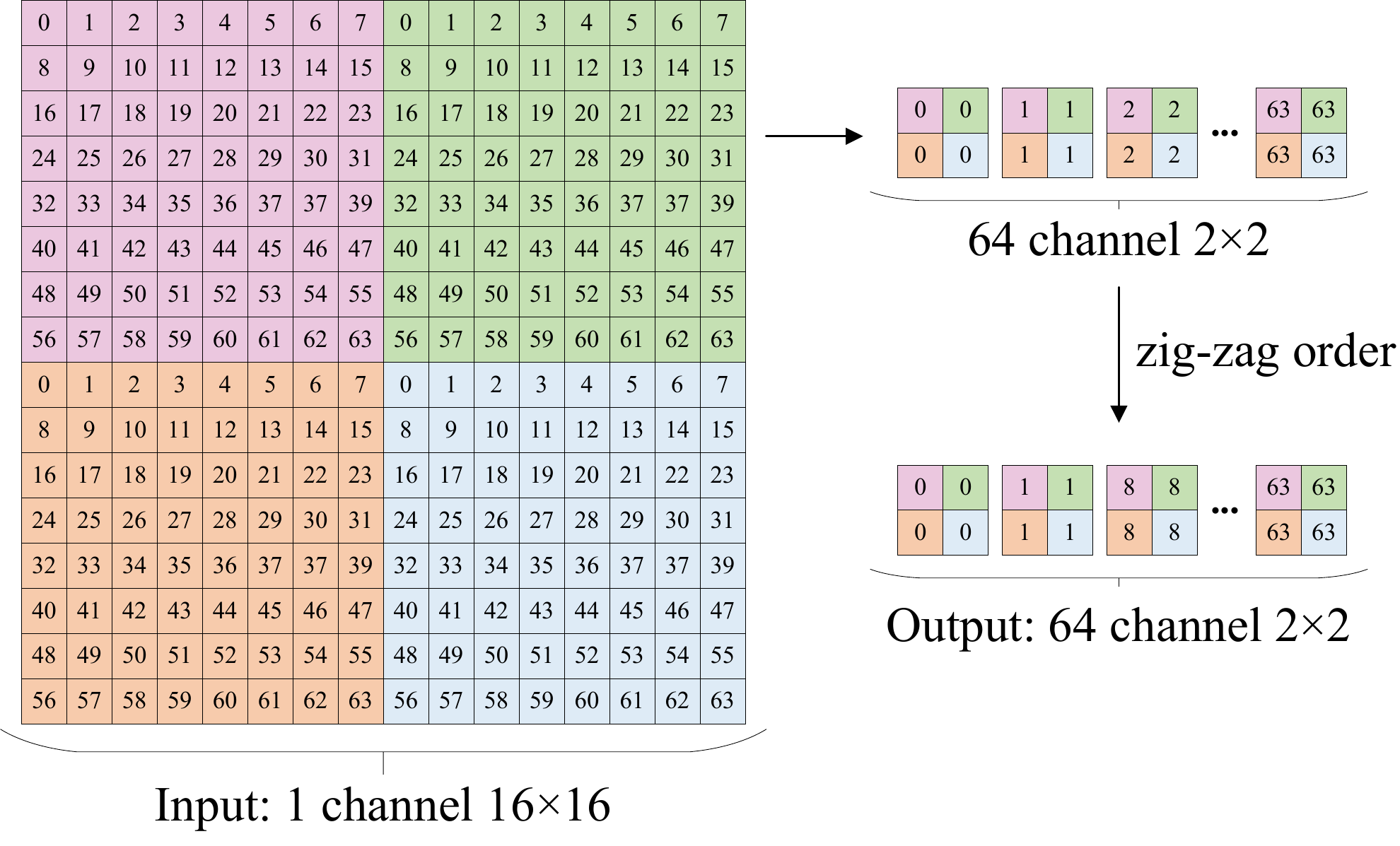}
  \caption{DCT coefficients rearrangement using a $ 16 \times 16 $ image as an example. Four $ 8 \times 8 $ DCT blocks are rearranged by frequency and zigzag scan.}
  \label{fig:arrange}
\end{figure}

JPEG encoder transforms the pixels in an $ 8 \times 8 $ block to a matrix of DCT coefficients in the same size and each coefficient in the matrix stands for one frequency. The top left corner of this matrix is the DC component, while the remaining 63 coefficients are AC components. As shown in \cref{fig:arrange}, we first adopt the same way as \cite{ehrlich2020quantization} to rearrange DCT coefficients so that the same frequency from all blocks are extracted together to form the spatial dimension, and different frequencies form the channel dimension. This operation converts Y, Cb, Cr components to 64 channels with $ \frac{1}{64}$ of their original spatial size. A lot of coefficients in AC components will go to zero after quantization. Therefore, we rearrange the channel dimension by zigzag scan in the original DCT matrix to make zero values as close as possible to exploit the structural information.  %, which help to obtain such structural dependent information in both spatial and channel dimensions that the entropy model can learn better. % Moreover, in order to entropy enhancement model of Y component to learn better, channel dimensions are further processed in reverse order.

\begin{figure}[htb]
  \centering
  \begin{subfigure}{0.3\linewidth}
      \centering
      \includegraphics[height = 4cm]{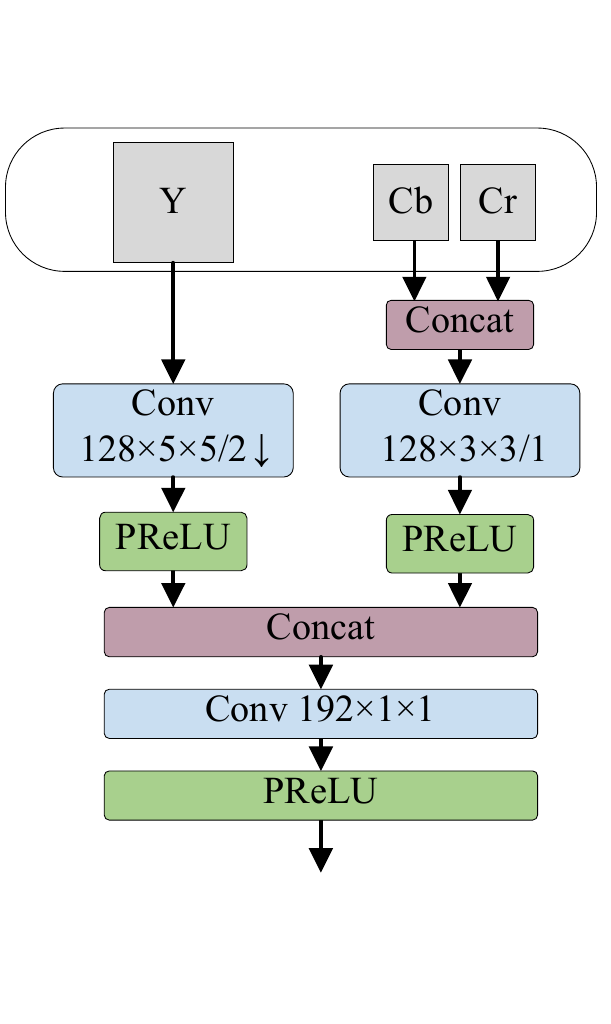}
      \subcaption{CFM}
      \label{fig:cross-yuv:cfm}
  \end{subfigure}
   \begin{subfigure}{0.34\linewidth}
      \centering
      \includegraphics[height = 4cm]{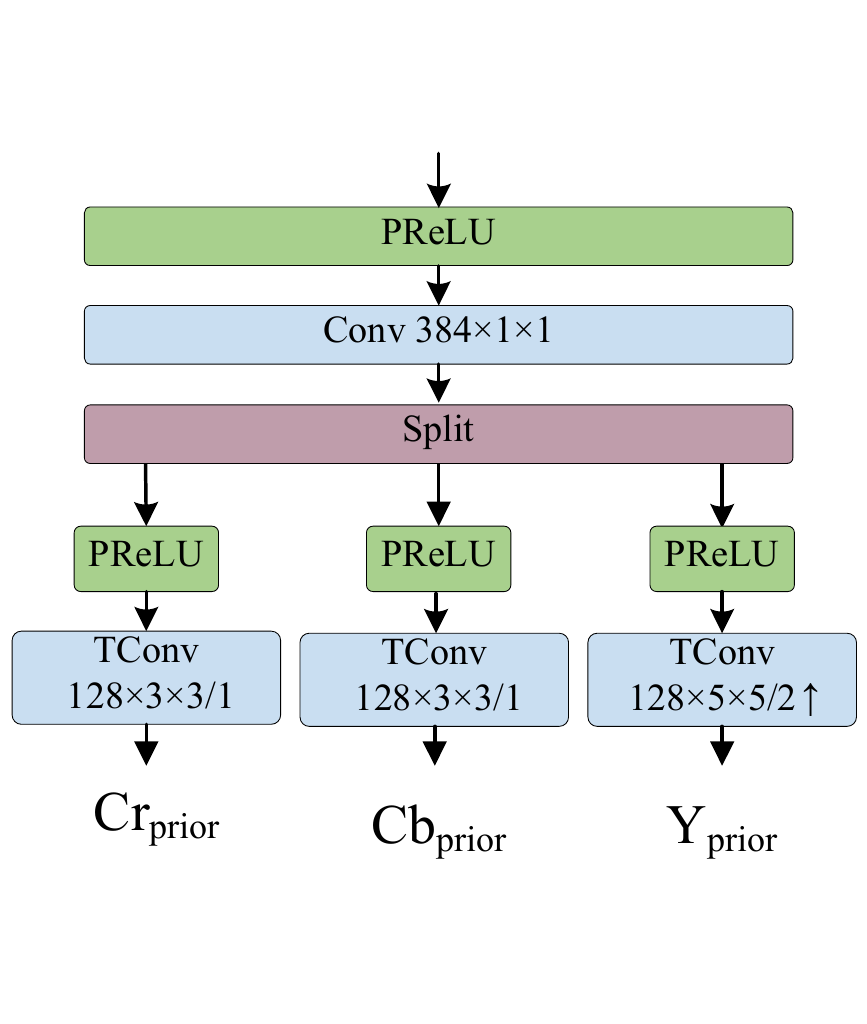}
      \subcaption{CPSM}
      	\label{fig:cross-yuv:cpsm}
  \end{subfigure}
  \begin{subfigure}{0.34\linewidth}
      \centering
      \includegraphics[height = 4cm]{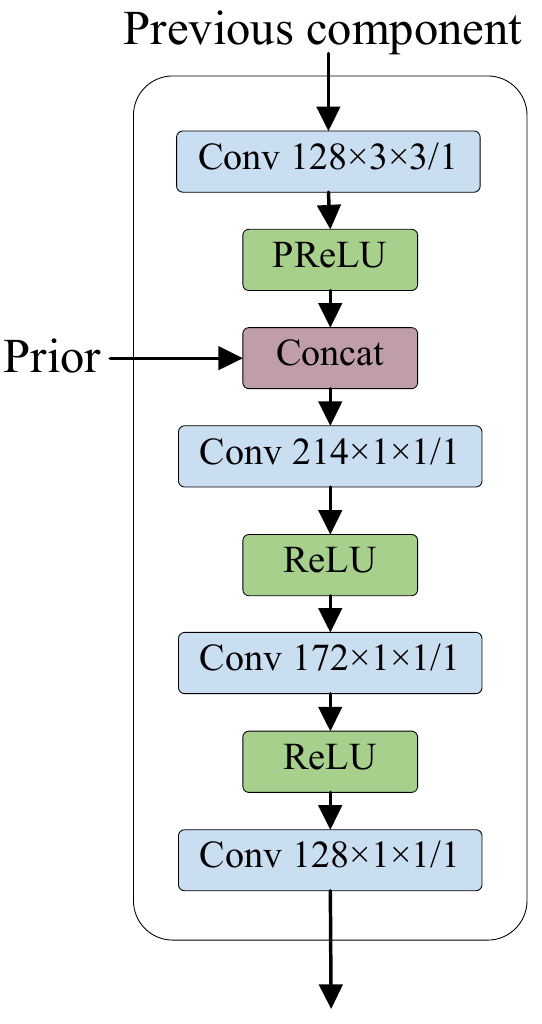}
      \subcaption{Entropy Parameters}
      \label{fig:cross-yuv:ep}
  \end{subfigure}
  \caption{Detailed architecture of CFM, CPSM and Entropy Parameters network.}
  \label{fig:cross-yuv}
\end{figure}

\subsection{Cross-Color Entropy Model}
\label{ssec:cem}
Cross-color correlation can be modeled both implicitly (through shared hyperprior) and explicitly (through the entropy parameters network).

 \emph{Hyperprior network} proposed in \cite{balle2018variational} can be viewed as an efficient entropy model which generates hyperprior $ \widetilde{z}$ as side information and then produces the scale parameters for Gaussian distribution conditioned on $ \widetilde{z}$. This method is improved in their later work \cite{minnen2018joint}, where hyperprior is combined with context-based predictions.
The same hyper-network as \cite{minnen2018joint} is used in our method to extract hyperprior from fused color components, which serves as side information and models the cross-color correlation implicitly.
However, this VAE-like model is unable to support YCbCr 4:2:0 format directly due to different spatial resolutions. Similar to \cite{egilmez2021transform}, we add Coefficient Fusion Model (CFM) and Coefficient Prior Split Model (CPSM) to \emph{Hyper Encoder} and \emph{Hyper Decoder} respectively. Architecture of CFM is shown in \cref{fig:cross-yuv:cfm}, through which the three color components are reshaped and fused. As shown in \cref{fig:cross-yuv:cpsm}, CPSM is used to split the prior of the three color components, producing $Y_{prior}$, $Cb_{prior}$ and $Cr_{prior}$.

%Y component is downsampled using a $ 5 \times 5 $ stride-2 convolution, whereas Cb and Cr components are concatenated and then processed using  a $ 3 \times 3 $ convolution without any downsampling to align the spatial shape. Then non-linear activated information from all color components are fused by concatenation and transformed by $ 1 \times 1 $ convolution with another non-linear activation. The output of CFM will be fed to  \emph{Hyper Encoder}. As shown in \cref{fig:cross-yuv:cpsm}, CPSM is an inverse process of CFM which is used to split the prior of the three color components, producing $Y_{prior}$, $Cb_{prior}$ and $Cr_{prior}$. 
 
 Each element of DCT coefficients is modeled as a single Laplace distribution with its own scale $ b $ and location $\mu$ parameters. We split $Cr_{prior}$ into $b_{cr }$ and $\mu_{cr}$ to obtain Laplace parameters of Cr component. $b_{cr}$ and $\mu_{cr}$ have the same size as Cr component. As formulated in \cref{eq:eqv}, the probability of Cr given $\widetilde{z}$ is calculated in a factorized manner. Subsequently, Cr component is fed to Entropy Parameters network (\cref{fig:cross-yuv:ep}) as context of Cb component and is fused with $Cb_{prior}$. The output of this model is split into $b_{cb} $ and $\mu_{cb}$. As a result, the probability mass function (PMF) of Cb component will be conditioned on both Cr component and hyperprior $ \widetilde{z}$ and is shown in \cref{eq:equ}.
 
 Cb and Cr components are upsampled by $ 3 \times 3 $ stride-2 transposed convolution and concatenated to serve as context of Y component. They are fed together with $Y_{prior}$ to Entropy Parameters network and we can obtain $ hyper_{y} $ (\cref{fig:frame}) to calculate the conditional distribution $p_{y |  \widetilde{z}, \coefcb, \coefcr}(y |  \widetilde{z}, \coefcb, \coefcr)$. Nevertheless, this PMF similar to Cr and Cb components is not powerful enough for the most informative Y component. In the following section, we propose a more suitable context modelling method to further reduce redundancy in Y component.
 %unlike from the above, the final Gaussian parameters of Y component applied in entropy coding is not acquired until further factorized and processed by MLCC in \cref{ssec:MLCC}.
 
 \begin{equation}
 \label{eq:eqv}
	\begin{aligned}
		 p_{\coefcr |  \widetilde{z}}(\coefcr |  \widetilde{z})  & = \prod^{N}_{i=0} p\left(  \coefcr_{i} |  \widetilde{z}\right)  \\
		 p(\coefcr_{i} |  \widetilde{z}) &= \int_{\coefcr_{i}-\frac{1}{2}}^{\coefcr_{i}+\frac{1}{2}} Laplace(\coefcr'|\mu_{\coefcr_{i}}, b_{\coefcr_{i}}) \,d\coefcr'
	\end{aligned}
\end{equation}
 \begin{equation}
 \label{eq:equ}
	\begin{aligned}
		 p_{\coefcb |  \widetilde{z}, \coefcr}(\coefcb |  \widetilde{z}, \coefcr)  &= \prod^{N}_{i=0} p\left(  \coefcb_{i} |  \widetilde{z}, \coefcr \right)  \\
		 p(\coefcb_{i} |  \widetilde{z}, \coefcr) & = \int_{\coefcb_{i}-\frac{1}{2}}^{\coefcb_{i}+\frac{1}{2}} Laplace(\coefcb'|\mu_{\coefcb_{i}}, b_{\coefcb_{i}}) \,d\coefcb'
	\end{aligned}
\end{equation}

\begin{figure*}[t]
  \centering
  \includegraphics[width=16cm]{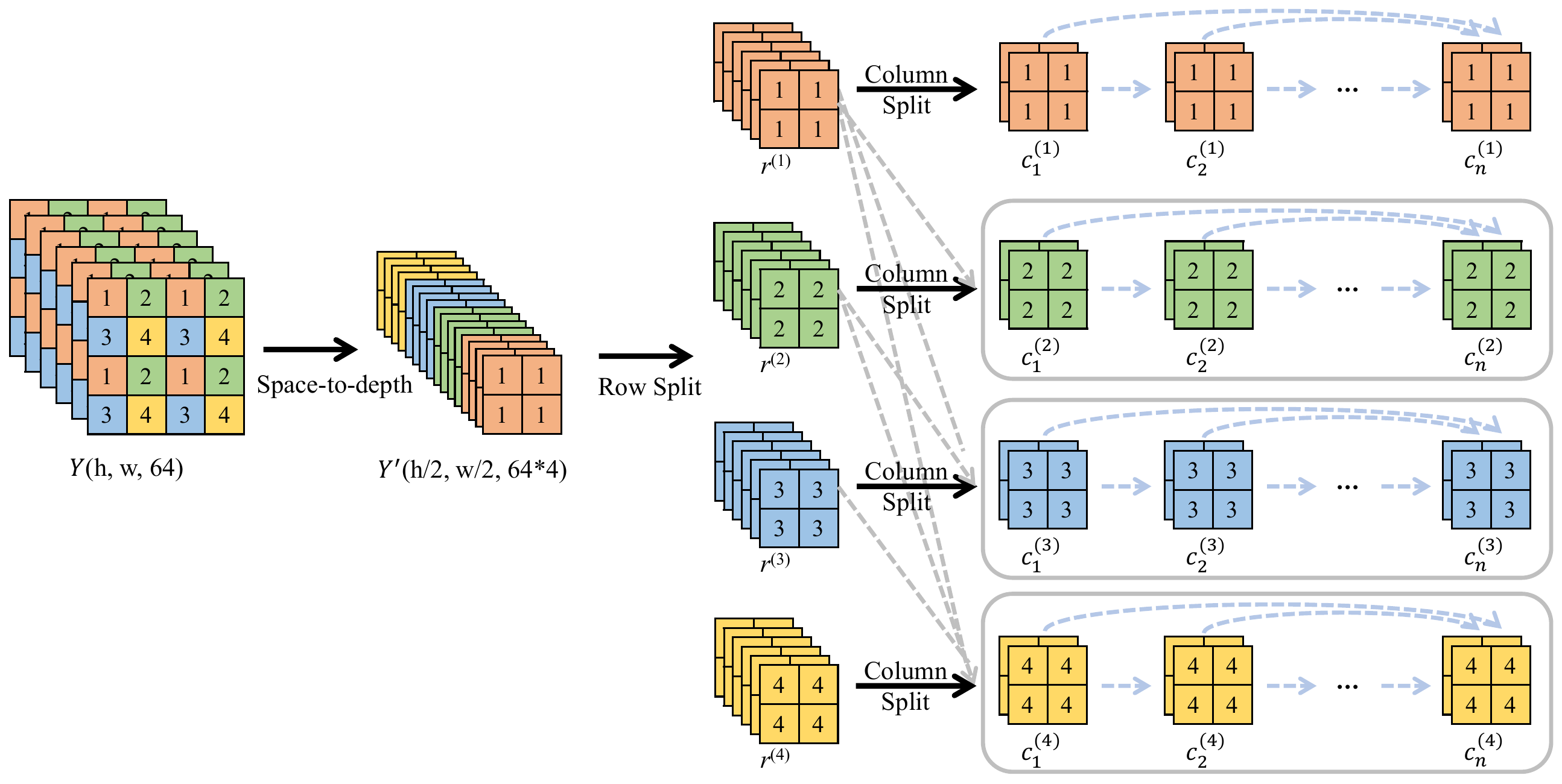}
  \caption{Matrix representation of our parallel context modelling method. Solid arrows indicate data operation and dotted arrows denote conditional relationships. Light grey and light blue dotted arrows align with Outer Channel and Inner Channel in \cref{fig:MLCC}, respectively. }
  \label{fig:matrix}
\end{figure*}

\subsection{Matrix Context Model}
\label{ssec:probabilistic}

Context modeling is an efficient technique to predict precise probabilistic distribution of unknown symbols based on adjacent symbols that have already been decoded. Previous learning-based codecs adopt a spatially auto-regressive context model, which requires decoding each symbol sequentially. While these methods are effective, they are impractical for real-world deployment due to low computational efficiency caused by the lack of parallelization \cite{minnen2018joint,johnston2019computationally}. Then, channel-conditional (CC) context models are explored in \cite{minnen2020channel}, which splits the symbol tensor along channel dimension into many equal-size slices and each slice can be modeled conditioned on all already decoded slices, providing much better parallelism. Subsequently, He \etal \cite{he2021checkerboard} proposes a novel spatial parallel context model with a two-pass decoding approach, where the symbol tensor is decomposed into two groups according to checkerboard pattern and then one group serves as context of the other to build conditional distribution. Meanwhile, a hierarchical probabilistic modeling idea, which is similar in spirit to spatially parallel context model, is prevalent in learned lossless image compression. Hierarchical models \cite{mentzer2019practical,cao2020lossless, zhang2020lossless} downsample an input image into different low-resolution representations and the probabilistic distribution of the input image is the product of the conditional distributions in multiple scales. %To further improve the precision of prediction, some hierarchical models (e.g.\  \cite{cao2020lossless, zhang2020lossless}) decompose the pixels at each scale into groups and each group is modeled conditioned on all previously decoded groups and scales.

In this paper, we propose a novel parallelizable matrix context model to enhance the entropy estimation of Y component. As shown in \cref{fig:matrix}, we first use space-to-depth operation (\ie inverse operation of PixelShuffle \cite{shi2016real}) to convert Y component  to  $ Y' $ ($ 64 \times 4 $ channels with $ \frac{1}{4}$ of the original spatial size). Then we partition $ Y' $ along channel dimension into 4 equal-size slices (\ie 4 rows in the matrix representation, $Y' = {r^{(1)} \bigcup r^{(2)} \bigcup r^{(3)} \bigcup r^{(4)} }$), where each row is modeled  conditioned on all previously decoded rows. Therefore, the conditional distribution $ p_{y|\widetilde{z}, \coefcb, \coefcr}(y |\widetilde{z}, \coefcb, \coefcr) $ can be calculated as:

\begin{equation}
	\label{eq:outerChannel}
	\begin{aligned}
		   & p_{y|\widetilde{z}, \coefcb, \coefcr}(y|\widetilde{z}, \coefcb, \coefcr) \\
		= & p_{y |  \widetilde{z}, \coefcb, \coefcr}(r^{(1)}, r^{(2)}, r^{(3)}, r^{(4)} |  \widetilde{z}, \coefcb, \coefcr) \\
		= & \prod^{4}_{i=1} p\left(  r^{(i)}|r^{(i-1)}, \cdots, r^{(1)}, \widetilde{z}, \coefcb, \coefcr\right)
	\end{aligned}
\end{equation}

Each row has 64 channels and there exists considerable redundancy. %across. these channels attributed to zig-zag scan mentioned in \cref{ssec:rearrangement}. 
Hence, each row is further partitioned to explicitly exploit this channel-wise correlation, \ie $ r^{(i)} = c^{(i)}_{1} \bigcup c^{(i)}_{2} \bigcup \cdots \bigcup c^{(i)}_{n} $, where $ n $ is the number of split columns at row $ i $. Let $ R^{(i)} = \left\lbrace r^{(i-1)}, \cdots, r^{(1)}, \widetilde{z}, \coefcb, \coefcr \right\rbrace $ denote the context and prior for $ r^{(i)} $ (specifically, $ R^{(1)} = \left\lbrace \widetilde{z}, \coefcb, \coefcr \right\rbrace $), we can further factorize \cref{eq:outerChannel} based on

\begin{equation}
	\label{eq:InnerChannel}
	\begin{aligned}
		 p\left(  r^{(i)}|R^{(i)}\right)   = 
		 \prod^{n}_{j=1} p\left(  c^{(i)}_{j}|c^{(i)}_{j-1}, \cdots, c^{(i)}_{1}, R^{(i)}\right)
	\end{aligned}
\end{equation}
where  $ c^{(i)}_{j} $ is column $j$ at row $i$, $ n $ is the number of columns at row $i$, and $ i = 1, 2, 3, 4 $.

 Let $C^{(i)}_{j} = \left\lbrace c^{(i)}_{1}, \cdots, c^{(i)}_{j-1}, R^{(i)} \right\rbrace  $ denote the context and prior for column $j$ at row $i$ (specifically, $C^{(i)}_{1} = \left\lbrace R^{(i)} \right\rbrace $) and coefficients within a column are conditionally independent and estimated by single Laplace model in parallel, we can further factorize \cref{eq:InnerChannel} based on \cref{eq:y}. Laplace parameters are derived from $ C^{(i)}_{j}$ according to \cref{ssec:MLCC}.

\begin{equation}
	\label{eq:y}
	\begin{aligned}
		 & p\left(  c^{(i)}_{j}|C^{(i)}_{j}\right)   =
		 \prod^{m_{j}}_{k=1} p\left(  y^{(i)}_{jk}|C^{(i)}_{j}\right)  \\
		 & p\left(  y^{(i)}_{jk}|C^{(i)}_{j}\right)  = \int_{y^{(i)}_{jk}-\frac{1}{2}}^{y^{(i)}_{jk}+\frac{1}{2}} Laplace(y'|\mu_{y^{(i)}_{jk}}, b_{y^{(i)}_{jk}}) \,dy'
	\end{aligned}
\end{equation}
where $ y^{(i)}_{jk} $ is coefficient $ k $ in column $ j $ at row $ i $, $ m_{j} $ is the number of coefficients in column $ j $, $ i = 1, 2, 3, 4 $, $ j=1, 2, \cdots, n $, and  $ k=1, 2, \cdots, m_{j} $. %Note that all the coefficients $ y^{(i)}_{jk} $ in the same column $ j $ are processed in parallel due to conditional independency.

According to the rearrangement in \cref{ssec:rearrangement}, the 64 channels at each row in our matrix context model represent different frequency, where higher frequency has been quantized more aggressively and contains less information. %actually represent 64 quantized DCT coefficients of an $ 8 \times 8 $ block, where the farther the AC component is away from the DC component, the less information it contains.
Therefore, we reverse the channel order in each row when formulating the matrix context (\ie $c^{(i)}_{1}$ in \cref{fig:matrix} is the AC coefficients representing the highest frequency in $r^{(i)}$). Moreover, we design non-uniform partition for the column dimension to balance this information asymmetry. Specifically, we let the number of columns be n=9, the lengths of each column ($c^{(i)}_{j}, j=1,2,\cdots,9 $) are set as 28, 8, 7, 6, 5, 4, 3, 2 and 1 respectively. 

\begin{figure*}[t]
  \centering
  \includegraphics[width=0.98\linewidth]{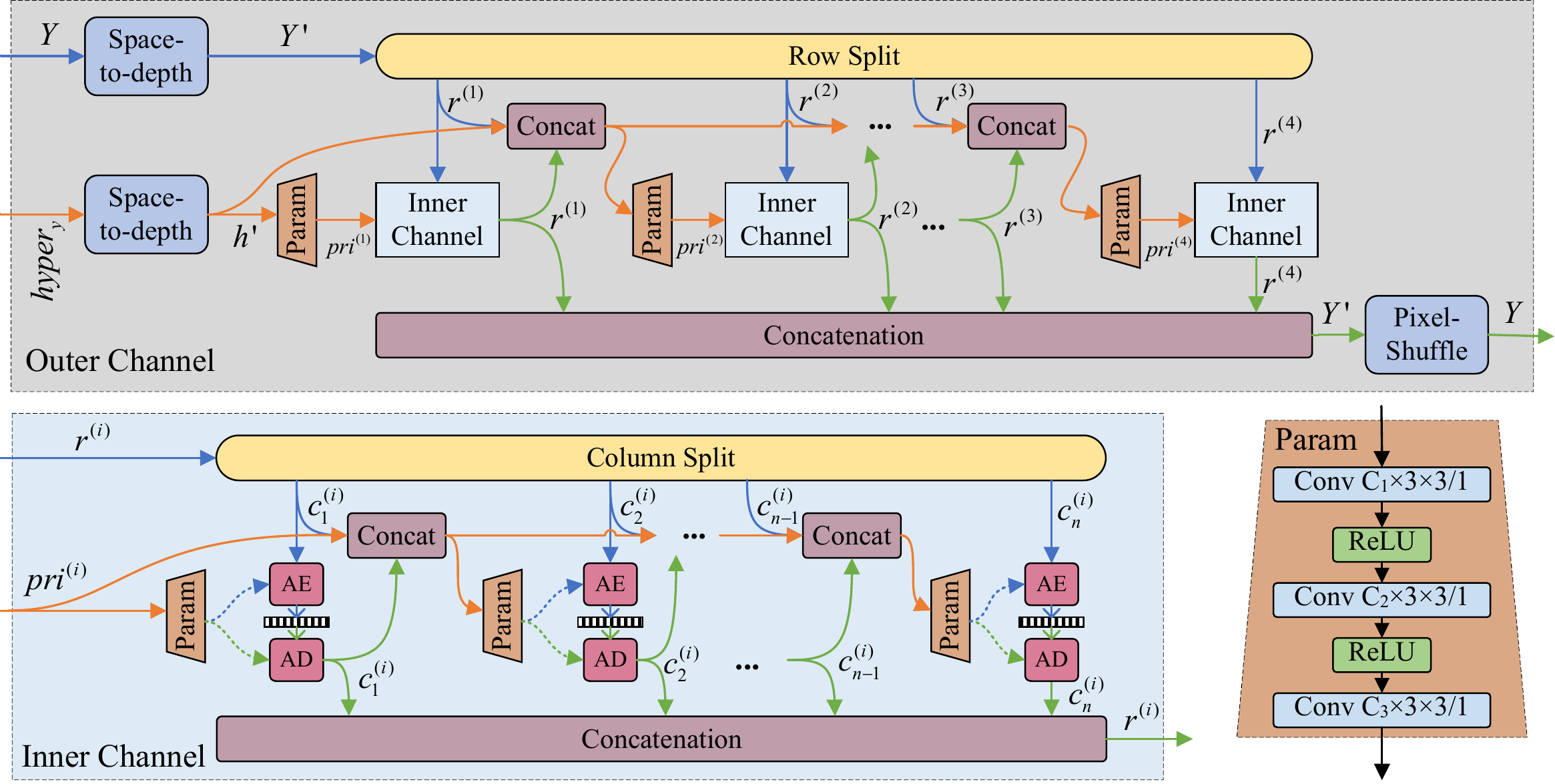}
  \caption{Detailed data-flow for applying multi-level cross-channel (MLCC) model. 
  Letting $n$, $m$ representing the channel number of input tensor and next slice to be modeled respectively, $C_1$, $C_2$ and $C_3$ are decided by 
  %$ C_1 $, $ C_2 $ and $ C_3 $ in \emph{Param} are set according to input tensor and next slice (
  $ C_1 = n - d, C_2 = n - 2*d, C_3=2*m, d = (n -2*m)//3$.
  %, $n$ and $m$ are the channel number of input tensor and next slice to \emph{Param}, respectively). 
  Blue and green lines indicate data-flow for encoding and decoding respectively, orange lines are shared.}
  \label{fig:MLCC}
\end{figure*}

\subsection{Multi-Level Cross-Channel Entropy Model}
\label{ssec:MLCC}
A deep neural network named Multi-Level Cross-Channel (MLCC) is designed to implement our matrix context entropy model to estimate Laplace parameters (location $ \mu $ and scale $ b $) in \cref{eq:y}, where we interpret the matrix context model as cross-channel autoregressive model: autoregression along the rows is viewed as Outer Channel (top in \cref{fig:MLCC}) to generate prior of each row (\ie $ pri^{(i)} $ in \cref{fig:MLCC}), autoregression along the columns in each row is modeled as Inner Channel (bottom left in \cref{fig:MLCC}) to generate Laplace parameters for each column. {MLCC} leverages both matrix context and $ hyper_{y} $ (\cref{ssec:cem}) to learn more powerful PMF for the most informative Y component. 

As shown in \cref{fig:MLCC}, We first adopt space-to-depth to rearrange $ hyper_{y} $ into $ h' $ as prior of Outer Channel ($ h' $ has $ 64 \times 4 $ channels with $ \frac{1}{4}$ spatial size of $hyper_{y}$). Meanwhile, Y component is reshaped by space-to-depth and then split into 4 rows (\cref{ssec:probabilistic}), the first row $ r^{(1)} $ is predicted conditioned solely on $ h' $.  However, unlike \cite{minnen2020channel}, this step in our method will generate prior $ pri^{(i)} $ for current row $ r^{(i)} $ rather than entropy parameters. Next, the current row and its own $ pri^{(i)} $ are sent to Inner Channel. In Inner Channel model, the row is partitioned into $ n $ columns (we set $ n $ to 9 in our method). And then the first column is compressed using a single Laplace entropy model with mean and scale conditioned only on $ pri^{(i)} $, while the entropy model for the remaining columns (\eg $ c^{(i)}_{j} $) are conditioned on $ pri^{(i)} $ and all decoded coefficients in previous columns (\eg $ \left\lbrace c^{(i)}_{1}, \cdots, c^{(i)}_{j-1} \right\rbrace $). After all columns in current row are decoded by Inner Channel model, they will be concatenated with $ h' $ and all previously decoded rows, and then processed by Outer Channel model again to generate prior $ pri^{(i+1)} $ for next row $ r^{(i+1)} $. Repeat this operation until all rows in Y component is encoded or decoded.

With this MLCC model, decoding will be slower than encoding because of the conditional relationship. In decoding stage those columns and rows must be decoded sequentially. 
However, all the coefficients in the same column can be processed in parallel, ensuring that the overall sequential complexity is constant (irrelevant of the input image size), which is $4 \times n$ (we let the number of columns be n=9). This guarantees that our method can be practical for recompressing high resolution JPEG images.%Particularly, during decoding, columns are first decoded sequentially and then concatenated as a row to decode next row, then all decoded rows are used to help decode all columns of next rows until end decoding. However, there are no major limitation for encoding, as all probability estimates are known ahead of time. 
\subsection{Loss Function}
\label{ssec:loss}
The expected code length arithmetic coding \cite{10.1145/214762.214771} can achieve, using our learned distribution as its probability model, is given by the cross entropy:
\begin{equation}
	\label{eq:loss}
	\begin{aligned}
		R  &= \mathbb{E}_{\widetilde{z} \sim  \widetilde{p}_{\widetilde{z}}}\left[ -\log_2 p_{\widetilde{z}}(\widetilde{z})\right]  
			 +  \mathbb{E}_{\coefcr \sim  \widetilde{p}_{\coefcr |  \widetilde{z}}}\left[ -\log_2 p_{\coefcr |  \widetilde{z}}(\coefcr |  \widetilde{z}) \right]   \\
		     &+ \mathbb{E}_{\coefcb \sim  \widetilde{p}_{\coefcb |  \widetilde{z}, \coefcr}}\left[ -\log_2 p_{\coefcb |  \widetilde{z}, \coefcr}(\coefcb |  \widetilde{z}, \coefcr) \right] \\
			 &+ \mathbb{E}_{y \sim  \widetilde{p}_{y |  \widetilde{z}, \coefcb, \coefcr}}\left[ -\log_2 p_{y |  \widetilde{z}, \coefcb, \coefcr}(y |  \widetilde{z}, \coefcb, \coefcr) \right]
	\end{aligned}
\end{equation}
where $ \widetilde{p} $ is the true distribution of DCT coefficients, $ p $ is estimated by entropy model. Our model is trained to minimize this cross entropy to  minimize the bit-length.

\section{Experiments}
\label{sec:exp}

\subsection{Settings}
\label{ssec:set}
\textbf{Datasets.} The training dataset comprises the largest 8000 images chosen from the ImageNet \cite{deng2009imagenet} validation set, where each image contains more than one million pixels. Similar to \cite{balle2016end, balle2018variational, he2021checkerboard}, each image is disturbed by uniform noise and downsampled. We evaluate our model on four datasets: \textbf{Kodak} \cite{kodak} dataset with 24 images, 100 images chosen from \textbf{DIV2K} \cite{agustsson2017ntire}, \textbf{CLIC} \cite{clic} \textbf{professional} test dataset with 250 images and \textbf{CLIC mobile} dataset with 178 images. Since our method processes images entirely in DCT domain, before fed to model, we use \textit{torchjpeg.codec.quantize\_at\_quality} \cite{torchjpeg} to extract quantized DCT coefficients with given JPEG quality level, which guarantees that the result is the same as using JPEG images generated from image libraries like Pillow. We fix the quality level of the training dataset at $ 75 $ if not specified. %which performs well on recompressing JPEG images of all quality levels except 95. For compressing JPEG images at quality level 95, we set the quality level of training dataset as 95. 

\textbf{Implementation details.} During training, $ 256 \times 256 $ pixel patches are randomly cropped from training data and then quantized DCT coefficients are extracted. Our model is implemented in PyTorch \cite{paszke2019pytorch} and we adopt Adam optimizer. The batch-size is $ 16 $ and the learning-rate is $ 10^{-4 }$. We apply gradient clipping for the sake of stability and train the model for $ 2000 $ epochs. All the speed testing results are obtained on single Nvidia GeForce GTX 1060 6GB (GPU) for learned methods and Intel(R) Xeon(R) CPU E5-2620 v4 @ 2.10GHz (CPU) for non-learned ones.

\subsection{Performance}
\label{ssec:comparison}
\textbf{Performance comparison with other JPEG recompression methods.} We compare the proposed model against other state-of-the-art methods for JPEG recompression on four test datasets mentioned in \cref{ssec:set}. We adopt our best model with $ 9 $ non-uniform channel slices, where the number of channels is split as $ [28, 8, 7, 6, 5, 4, 3, 2, 1] $.
As shown in \cref{tab:comparison:set}, with quality level set as 75, our method achieves lowest bit rate on all evaluation datasets and obtains about $ 30\% $ compression savings. 
%Moreover, we compare the encoding and decoding time of our method with other JPEG recompression methods. As shown in the right of \cref{tab:comparison:set},
Our method is much faster than CMIX but slower than JPEG XL and Lepton. However, it is worth noticing that our implementation of arithmetic coder is naive and our model has not been optimized to achieve the fastest speed. 

\begin{figure}[htb]
  \centering
  \includegraphics[width=.45\textwidth]{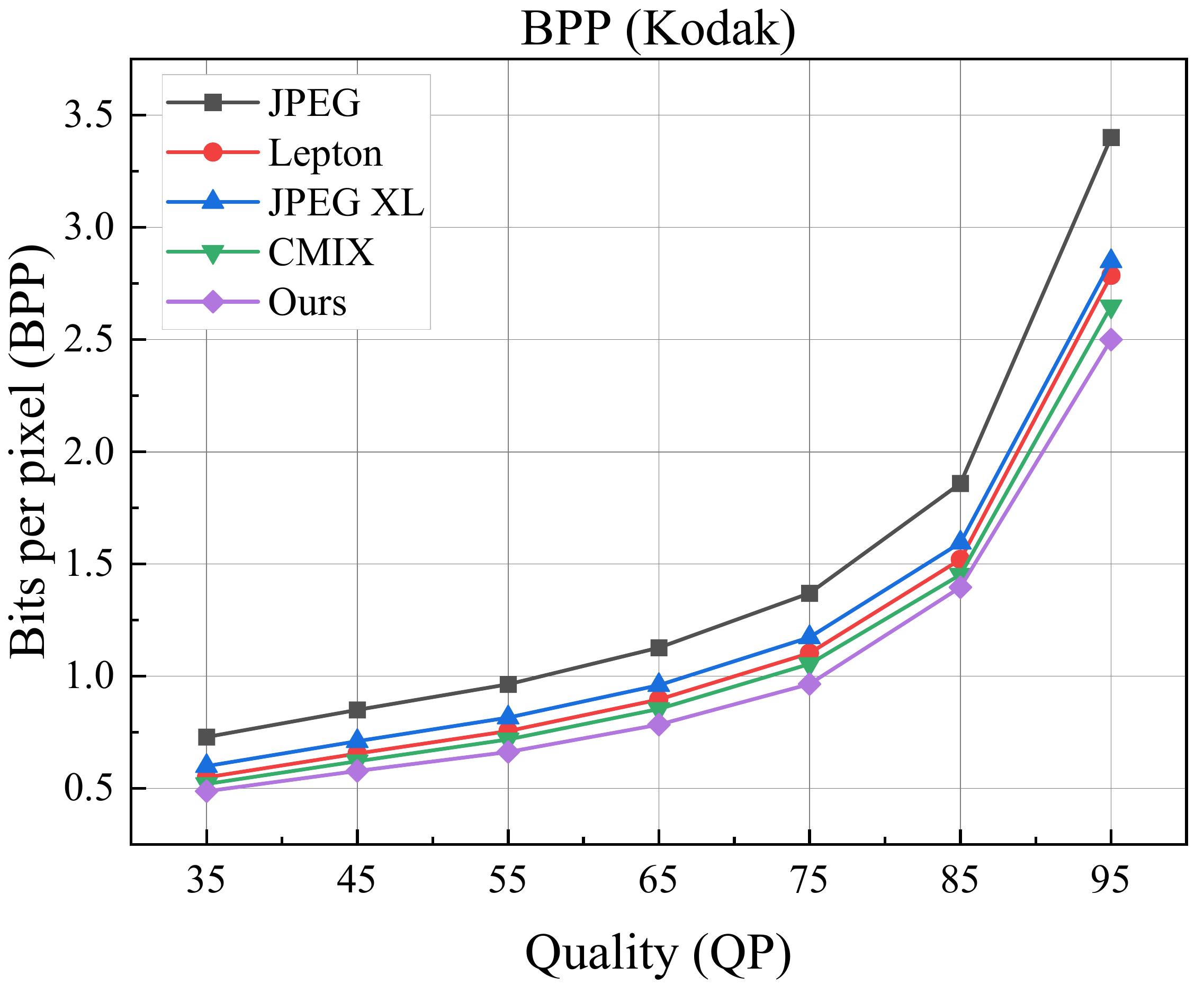}
  \caption{Comparison of bits per pixel (BPP) on Kodak dataset when recompressing JPEG images of different quality levels (QP). At QP 95, we use the model trained with QP 95. At other points, we use the model trained with QP 75.}
  \label{fig:comparison:qp}
\end{figure}

\textbf{Performance on different quality levels.} We test our models on Kodak with $ 7 $ different JPEG quality levels (\ie $ quality=35, 45, 55, 65, 75, 85, 95 $).
The results are presented in \cref{fig:comparison:qp}. It shows that our method still outperforms other methods, which shows that our model trained for $quality=75 $ can generalize well to different quality levels except very high quality like 95. More detailed results are given in the appendix. %However, when $ quality > 75 $, the performance of our model drops significantly, and the improvement over lepton becomes smaller, which implies that the model trained using $ quality = 75 $ does not generalize well to $ quality = 85, 95 $.  %reflects the instability of our model caused by the difference between training datasets ($ quality = 75 $) and testing datasets.
%Note that popular non-learned methods like lepton and JPEG XL also has similar problem, i.e., their performance drops as the quality level increases.

\textbf{Performance comparison with other learned lossless compression methods.} We compare our method with representative learned lossless image compression methods designed for PNG images, including IDF \cite{hoogeboom2019integer} and multi-scale model \cite{zhang2020lossless}. These methods are designed for RGB 4:4:4 format, so we convert JPEG 4:2:0 input data to RGB 4:4:4 format by upsampling Cb and Cr components. This upsampling operation increases resolution and may cause unfair comparison. Consequently, we also carry out experiments with JPEG 4:4:4 source format and convert it to RGB 4:4:4 as model input. By modifying these models slightly, these methods can also deal with JPEG 4:2:0 format directly, we present results of this kind of experiments in the appendix. 

\begin{figure}[htb]
  \centering
  \includegraphics[width=.43\textwidth]{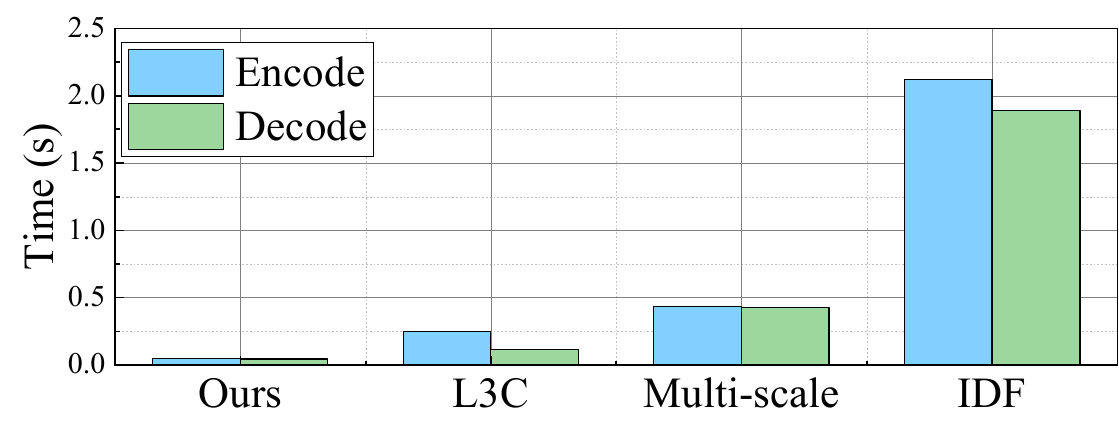}
  \caption{Time spent on neural networks compared with representative learned lossless compression methods. These models are evaluated on Kodak in JPEG 4:4:4 with QP 75.}
  \label{fig:comparison:net-speed}
\end{figure}

As shown in \cref{tab:comparison:dct}, both for JPEG 4:2:0 and JPEG 4:4:4 format, our models outperform IDF and multi-scale model by a large margin. %Even for models directly trained with DCT 4:2:0 and DCT 4:4:4 format, our method still performs better, which strongly proves the superiority of our method.
Additionally, we evaluate neural network latency of our model, L3C \cite{mentzer2019practical}, IDF \cite{hoogeboom2019integer} and multi-scale \cite{zhang2020lossless} in \cref{fig:comparison:net-speed}, which shows our model is faster.
% todo: what modification
% todo: 444 related changes
% todo: network speed compare

\begin{table*}[]
    \centering
    \begin{tabular}{c|p{2.4cm}p{2.4cm}p{2.4cm}p{2.4cm}|cc}
    	\hline
    	~ & \multicolumn{4}{c|}{BPP and Savings (\%)} & \multicolumn{2}{c}{time (s)}\\
    	\hline
        Method & Kodak & DIV2K  & CLIC.mobile & CLIC.pro & Encoding & Decoding\\
         \hline
         JPEG \cite{wallace1992jpeg}            
         & 1.369    &   1.285  &  1.099  & 0.922  &-&-  \\
         Lepton \cite{horn2017design}
         & 1.102 \textcolor[RGB]{0,180,0}{$ (19.50\%) $} &1.017 \textcolor[RGB]{0,180,0}{$ (20.86\%) $}& 0.863  \textcolor[RGB]{0,180,0}{$ (21.47\%) $} & 0.701  \textcolor[RGB]{0,180,0}{$ (23.97\%) $} &0.239&0.127\\
         JPEG XL \cite{alakuijala2019jpeg, alakuijala2020benchmarking}
         & 1.173 \textcolor[RGB]{0,180,0}{$ (16.70\%) $} &1.072 \textcolor[RGB]{0,180,0}{$ (16.58\%) $}& 0.908
         \textcolor[RGB]{0,180,0}{$ (17.38\%) $} & 0.744  \textcolor[RGB]{0,180,0}{$ (19.30\%) $}
         &\textbf{0.179}&\textbf{0.125}\\
         CMIX \cite{cmix}
         & 1.054 \textcolor[RGB]{0,180,0}{$ (23.01\%) $} &0.931 \textcolor[RGB]{0,180,0}{$ (27.55\%) $}& 0.804
         \textcolor[RGB]{0,180,0}{$ (26.84\%) $} & 0.648  \textcolor[RGB]{0,180,0}{$ (29.72\%) $}
         &152.9&154.5\\
         Ours  & \textbf{0.965} \textcolor[RGB]{0,180,0}{\bm{$(29.51\%)$}}& \textbf{0.892} \textcolor[RGB]{0,180,0}{\bm{$(30.58\%) $}} & \textbf{0.772} \textcolor[RGB]{0,180,0}{\bm{$(29.75\%)$}} & \textbf{0.624} \textcolor[RGB]{0,180,0}{\bm{$(32.32\%)$}}
         &1.131 &1.023\\
        \hline
    \end{tabular}
    \caption{ Performance comparison on various datasets. The encoding and decoding time are evaluated on Kodak with QP 75.
    }
    \label{tab:comparison:set}
\end{table*}

\begin{table}[]
    \centering
    \begin{tabular}{c|l|c|c}
    		\hline
        Source format & Method & Input format & BPP \\
         \hline
         \multirow{3}{*}{JPEG 4:2:0}
         & Multi-scale \cite{zhang2020lossless} & RGB 4:4:4  &  4.398    \\
         & IDF \cite{hoogeboom2019integer} & RGB 4:4:4 &  6.964 \\
         & Ours & DCT 4:2:0 &  \textbf{0.965} \\
         \hline
         \multirow{3}{*}{JPEG 4:4:4}
         & Multi-scale \cite{zhang2020lossless} & RGB 4:4:4 &  4.604    \\
         & IDF \cite{hoogeboom2019integer} & RGB 4:4:4 &  7.059 \\
         & Ours & DCT 4:4:4 &  \textbf{1.122} \\
         \hline
    \end{tabular}
    \caption{Performance comparison with learned lossless compression methods.
    }
    \label{tab:comparison:dct}
\end{table}

\subsection{Ablation study}
\label{ssec:ablation}
We test a serial of models on Kodak with quality level 75 to investigate the effect of cross-color entropy model, multi-level cross-channel entropy enhancement model and non-uniform channel slices.

% \textbf{DCT domain \textit{vs.} pixel domain. }There are two models to verify the importance of deploying model in the DCT domain: \textit{multi-scale + pixel domain} operate in pixel domain with this architecture that has same multi-scale entropy model in proposed MLCC but has no channel model, since pixel domain has only $ 3 $ channels. In additional, the last representation in multi-scale is estimated by non-parametric piecewise linear density model proposed in \cite{balle2016end}. \textit{multi-scale + DCT domain} operate entirely in the DCT domain. It adopts same framework with proposed method but drop channel model from MLCC to compare fairly. Moreover, in order to the results looks more objective, we add a item which is the rate after converting JPEG to PNG. As shown in \cref{tab:ablation}, the multi-scale model operated in pixel domain is effective for the PNG format compression, but it is invalid for the JPEG image, which demonstrates that it is necessary to deploy model in the DCT domain.

\begin{table}[]
    \centering
     \begin{tabular}{l|c|c|c}
     	\hline
         Method &Parameters & BPP & Savings \\
         \hline
%         JPEG (original)          & 1.369 & \\
%         Lepton                   & 1.102 & 19.50\% \\
%         JPEG XL                  & 1.173 & 16.70\% \\
%         CMIX                     & 1.054 & 23.01\% \\
         Ours                     &32.3M&\textbf{0.965}& \textbf{29.51\%} \\
 %        \hline
 %        \multicolumn{3}{l}{Effectiveness of cross-color entropy model} \\
         \hline
         Cross-color case 1        & 36.9M& 0.983 & 28.20\%   \\
         Cross-color case 2        & 31.5M& 0.973 & 28.93\%   \\
         Cross-color case 3        & 88.0M& 0.968 & 29.29\%   \\
         \hline
%         \multicolumn{3}{l}{Effectiveness of MLCC model} \\
%         \hline
         Only Outer Channel        &22.5M& 1.027 & 24.98\%   \\
         Only Inner Channel        &9.1M& 1.012 & 26.08\%   \\
         Column-to-row            &13.0M& 0.988 & 27.83\%   \\
         \hline
%         \multicolumn{3}{l}{Effectiveness  of  non-uniform  slice} \\
%         \hline
         Uniform 8 slices         &30.0M& 0.986 & 27.98\%   \\
         Non-uniform 8 slices     &31.5M& 0.966 & 29.44\%   \\
         \hline
    \end{tabular}
    \caption{Ablation study.}
    \label{tab:ablation}
\end{table}

\textbf{Effectiveness of cross-color entropy model.} To verify the effectiveness of our cross-color entropy model, we compare three models. \textbf{Cross-color case 1:} Y, Cb, and Cr components are modeled totally independent of each other, \ie there are three side information $ \widetilde{z}_{Y},\widetilde{z}_{Cb},\widetilde{z}_{Cr}$, and Y (with MLCC), Cb and Cr components are dependent on $\widetilde{z}_{Y},\widetilde{z}_{Cb},\widetilde{z}_{Cr}$ respectively. \textbf{Cross-color case 2:} Hyperprior $ \widetilde{z}$ is shared by Y, Cb, and Cr components, and these three color components are conditioned on the shared $ \widetilde{z}$ except that the entropy model of Y component is enhanced by MLCC. Here $ \widetilde{z}$ provides implicit cross-color correlation while no explicit modeling is used. \textbf{Cross-color case 3:} Three color components are modeled totally independent of each other, while they are treated equally and are all modeled by independent hyperprior with corresponding MLCC module. We give detailed architectures in the appendix. %TODO: draw aichitecture in sup. %and the same architecture as that for Y component is applied on Cr and Cb except cross-color dependency. %and contains three separate network which corresponds three components respectively. Each separate network in this model comprise same hyperprior network and MLCC as proposed model, where single component is fed to hyperprior network and then the output of hyperprior network decoder is sent to MLCC as hyper prior.  
As shown in \cref{tab:ablation}, savings of case 1 and case 2 are lower than our proposed model and case 2 is better than case 1, indicating that both implicit and explicit cross-color correlation modeling is helpful to bit saving. Although the network capacity and parameter number are the largest in case 3, its performance is still slightly worse than our model, proving the effectiveness of our cross-color entropy model.
%Moreover, \cref{tab:ablation:ycbcrz} illustrate Cb, Cr contains less information where minuscule bitrates can be achieved with simple entropy models, and Y component is the key factor affecting compression performance. Fortunately, cross-color entropy model exploit shared side information and cross-color dependency to improve the accuracy of PDF on Y component, resulting in better compression performance.

%\begin{figure}[htb]
%  \centering
%  \includegraphics[height=6cm]{latex/img/bit_analysis.pdf}
%  \caption{The distribution of bit rate for our proposed %model.}
%  \label{fig:comparison:c_entropy}
%\end{figure}

\textbf{Effectiveness of MLCC model.} %As shown in the left of \cref{fig:comparison:c_entropy}, Y component occupies much more bitrates than Cb and Cr components.  %thus it is reasonable to add more powerful PDF for the most informative Y component. Subsequently, 
To verify the effectiveness of MLCC, we replace MLCC with three different models while keeping other parts of the model unchanged. \textbf{Only Outer Channel} drops the Inner Channel module in MLCC, which means there are no column split operation. \textbf{Only Inner Channel} has no space-to-depth and row split operations, which only has Inner Channel module in MLCC. \textbf{Column-to-row} is a variation of our row-to-column MLCC, which adopts column split first and then row split. Details about these models are given in the appendix. Shown in \cref{tab:ablation}, the above three replacements of MLCC deteriorate the compression savings, which verifies the effectiveness of MLCC.

\textbf{Effectiveness of non-uniform slices.} %As shown in the right of \cref{fig:comparison:c_entropy}, we analyze the entropy of each channel in Y component, and find that channels with smaller indexes (corresponding to higher frequency) occupy less entropy, which demonstrates the information asymmetry mentioned in \cref{ssec:probabilistic}. 
We compare two models to verify that non-uniform column slicing in MLCC is more effective for JPEG recompression. \textbf{Uniform 8 slices} divides rows evenly into $ 8 $ columns, while \textbf{non-uniform 8 slices} divides rows into $ 8 $ columns of size $ [36, 7, 6, 5, 4, 3, 2 , 1] $ respectively. As demonstrated in \cref{tab:ablation}, \textbf{non-uniform 8 slices} model has the same column number as \textbf{uniform 8 slices} model but its compression ratio is about $ 1.5\% $ higher.

\section{Conclusion}
We propose a novel Multi-Level Cross-Channel entropy model for lossless recompression of existing JPEG images, which achieves state-of-the-art performance on Kodak, DIV2K, CLIC.mobile and CLIC.pro and has reasonable running speed. We also show that our method trained with quality level 75 can generalize well on other quality levels except very high quality like 95. To the best of our knowledge, this is the first learned method targeting lossless recompression of JPEG images. For future work, we will explore the generalizability on very high quality levels.

%%%%%%%%% REFERENCES
{\small
\bibliographystyle{ieee_fullname}
\bibliography{egbib}
}

\end{document}

\title{Supplementary Material}

%\and
\maketitle
\section{Analysis of of bit rate distribution}
\label{sec:analysis}

\begin{figure}[htb]
 \centering
 \includegraphics[height=6cm]{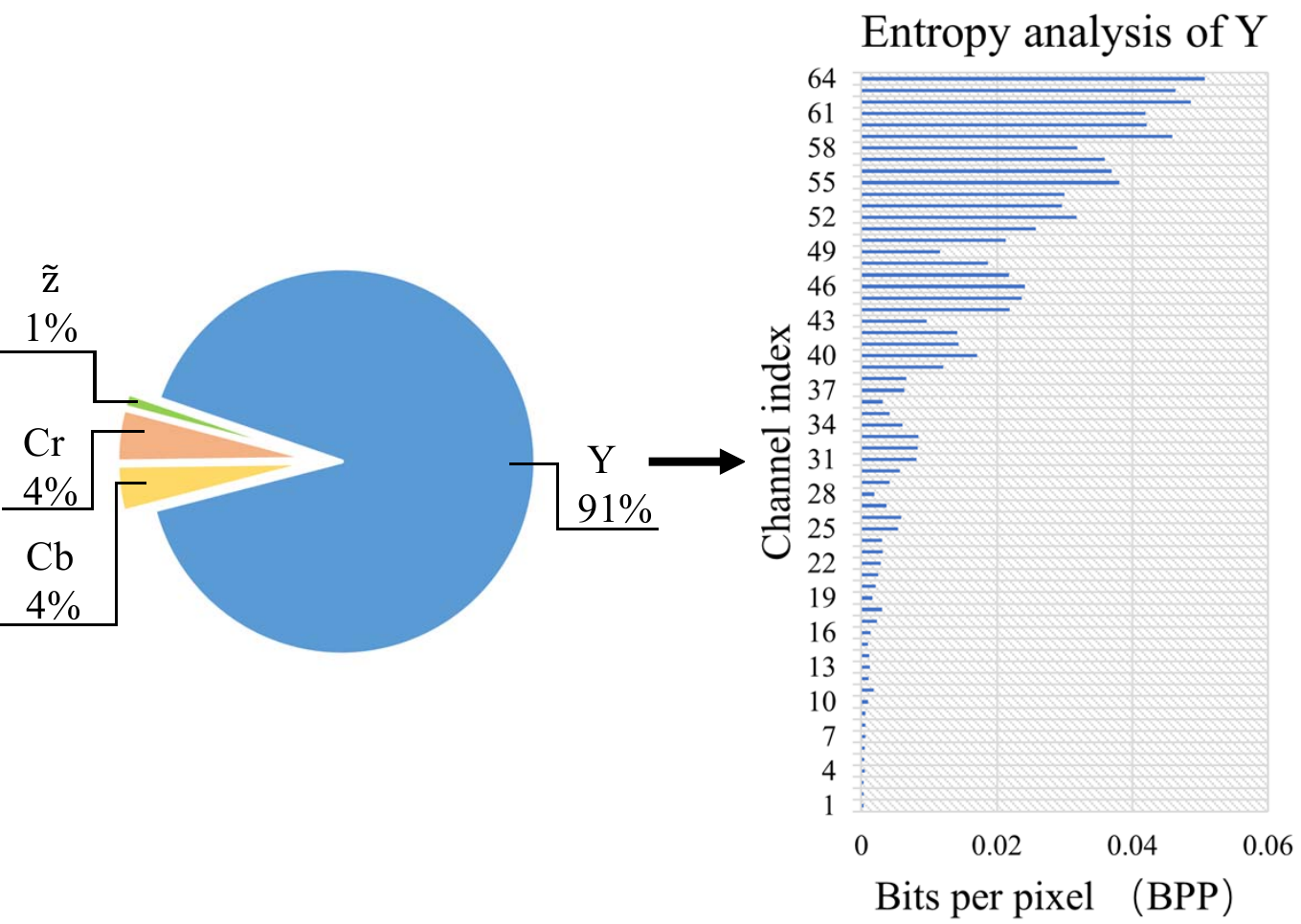}
 \caption{The distribution of bit rate for our proposed model.}
 \label{fig:comparison:c_entropy}
\end{figure}

As shown in the left of \cref{fig:comparison:c_entropy}, the Y component occupies much more bitrates than Cb and Cr components. Then we analyze the entropy of each channel in the Y component and find that channels with smaller indexes (corresponding to higher frequency) occupy less entropy, which demonstrates the information asymmetry mentioned in Sec.\ 3.4 in the main text.

\section{Entropy coding and probability model}
\label{ssec:entropy}
Lossless compression technique is essentially an entropy coder along with a probability model. JPEG algorithm adopts Huffman coding together with Huffman tables defining the probability model to compress quantized DCT coefficients losslessly. However, these coefficients are considered independently and identically distributed (i.i.d.) under this fixed probability model, resulting in a mismatch between estimated and actual data distribution, which decreases compression savings. 

Our method contains two significant improvements in this aspect. First, our method uses Laplace distribution with different parameters for each coding symbol to obtain an adaptive probability model, where the two parameters (scale $ b $ and location $ \mu $) of each Laplace distribution are learned by neural networks. It is known that AC coefficients of Fourier-related transformations, like DCT coefficients of JPEG algorithm, obey Laplace distribution \cite{10.1117/1.1344592}. Second, we replace the Huffman coding with arithmetic coding, which is a more efficient technique that almost achieves the lower bound entropy for long enough symbol streams. 
%todo Guassian experiment

\section{Details about performance on different quality levels}
\label{sec:diff_qp}
\begin{figure}[htb]
  \centering
  \includegraphics[width=.45\textwidth]{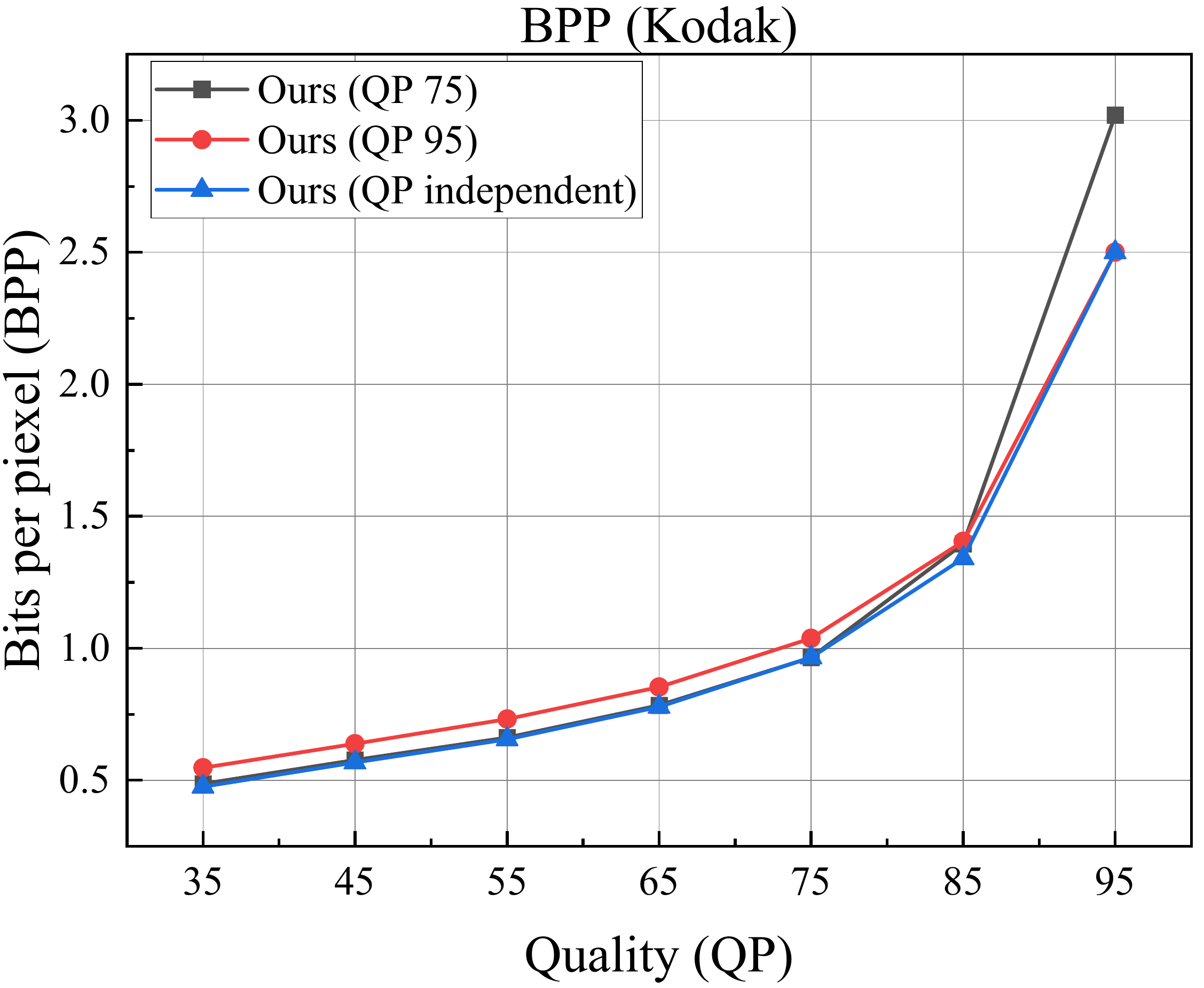}
  \caption{Comparison of bits per pixel (BPP) on Kodak dataset when recompressing JPEG images of different quality levels (QP).}
  \label{fig:qp}
\end{figure}

\cref{tab:QP} is the detailed data corresponding to Figure 6 in the main text. All results at row \textbf{Ours (QP 75)} and row \textbf{Ours (QP 95)} are obtained by our models trained with QP 75 and QP 95 respectively. While results in row \textbf{Ours (QP independent)} are generated by $7$ models trained with QP $35, 45, 55, 65, 75, 85$ and $95$ respectively to ensure that QP value is exactly the same for training and testing. Data of these three rows are visualized by \cref{fig:qp}. It shows that the \textbf{Ours (QP 75)} has comparable performance with \textbf{Ours (QP independent)} at all QPs lower than 85, which further demonstrates that our model trained for $quality=75 $ can generalize well to different quality levels except very high quality like 95.%It shows that our method still outperforms other methods, which shows that our model trained for $quality=75 $ can generalize well to different quality levels except very high quality like 95.
\begin{table*}[]
    \centering
    \begin{tabular}{c|p{1.5cm}p{1.5cm}p{1.5cm}p{1.5cm}p{1.5cm}p{1.5cm}p{1.5cm}}
    		\hline
        Method & QP35 & QP45  & QP55 & QP65 & QP75 &QP85 &QP95\\
         \hline
         JPEG             
         & 0.729    &   0.850  &  0.964  & 1.127  &1.369 &1.859 &3.401  \\
         Lepton 
         &0.549
         &0.655
         &0.755 
         &0.896 
         &1.102
         &1.520
         &2.786\\
         JPEG XL 
         &0.599
         &0.710
         &0.815
         &0.960
         &1.173
         &1.595
         &2.849\\
         CMIX 
         &0.519
         &0.621
         &0.718
         &0.853
         &1.054
         &1.452
         &2.648\\
         Ours (QP 75)  
         &0.487
         &0.577
         &0.662
         &0.784
         &0.965
         &1.396
         &3.022\\
         Ours (QP 95)  
         &0.547
         &0.638
         &0.732
         &0.853
         &1.038
         &1.405
         &2.500\\
         Ours (QP independent)
         &0.476
         &0.568
         &0.655
         &0.778
         &0.965
         &1.341
         &2.500\\
        \hline
    \end{tabular}
    \caption{Comparison of bits per pixel (BPP) on Kodak dataset when recompressing JPEG images of different quality levels (QP).
    }
    \label{tab:QP}
\end{table*}

% \section{More performance comparison}
% \label{sec:comparison}

\section{Details about comparison with learned lossless methods}
\label{sec:training}
We reproduce the multi-scale model following the instructions in the original paper, and our reproduced model achieves bits per sub-pixel (BPSP) of 3.942 (computed by negative log-likelihood) on ImageNet64 dataset, which is slightly better than BPSP 3.96 presented in the original paper. 

We also reproduce IDF based on the source code released by the authors. Nevertheless, the original IDF can only accept input with fixed size due to the logistic mixture model (LMM) used for latent prior $z_L$ ($L$ is the level of flows). To make IDF resolution-adaptive, we replace the LMM with the univariate non-parametric density model used in \cite{balle2018variational}. Our reproduced model achieves BPSP of 3.879 (computed by negative log-likelihood) on ImageNet64 dataset, slightly better than BPSP of 3.90 given in the original paper.

As stated in the main text, these two methods are designed for RGB 4:4:4 input, so we convert JPEG 4:2:0 input data to RGB 4:4:4 or YCbCr 4:4:4 by upsampling Cb and Cr components. This upsampling operation increases resolution and may cause unfair comparison. Consequently, we also carry out experiments with JPEG 4:4:4 source format and convert it to RGB 4:4:4, YCbCr 4:4:4 and DCT 4:4:4 as model input, which ensures a fair comparison. We make our method suitable for JPEG 4:4:4 images by removing the downsampling and upsampling of Y component in CFM and CPSM. The full experiment settings are given in~\cref{tab:lossless}. We use Pillow library to read RGB values from the source JPEG images, and YCbCr values are converted from RGB values.

%We consider two kinds of source JPEG to be compressed (JPEG 4:2:0 and JPEG 4:4:4) and carry out experiments on four input formats, \ie RGB 4:4:4, YCbCr 4:4:4, DCT 4:4:4 and DCT 4:2:0 as shown in ~\cref{tab:lossless}. We use Pillow library to read RGB values from the source JPEG images, and YCbCr values are converted from RGB values. 

For IDF and multi-scale model using DCT coefficients as input, we do not adopt the DCT coefficients rearrangement proposed in the main text to avoid enormous architecture modification (this is because they are originally designed for thin input format like RGB values with only 3 channels). To deal with DCT 4:2:0 input, we reshape Y component to 4 channels by space-to-depth operation and then concatenate these 4 channels with Cb and Cr components to form 6-channel inputs with $\frac{1}{4}$ of their original resolutions. Meanwhile, both of the methods need to be adjusted slightly to fit this kind of input. For multi-scale model, the number of input channels is changed from 3 to 6, and autoregression over RGB channels is disabled. For IDF, we start from the original IDF and increase the number of input channels to 6.

For all of these experiments, we use Adam optimizer and MultiStepLR scheduler. We use the same training set and use Kodak for evaluation, and the quality level is 75. Other settings and experiment results are presented in \cref{tab:lossless}. For both JPEG 4:2:0 and JPEG 4:4:4 images, our method outperforms other methods on all input formats by a large margin. It is worth noting that though for JPEG 4:4:4 images, our method for compressing Cb and Cr components is relatively simple, the performance is still superior. And it is obvious that lossless image compression methods designed for PNG do not perform well on recompressing JPEG 4:2:0 and JPEG 4:4:4 images.

\begin{table*}[]
    \centering
    \begin{tabular}{c|c|c|c|c|c|c|c|c|c}
    		\hline
        \tabincell{c}{Source\\format} & Model &\tabincell{c}{Input\\ format} & Crop size & \tabincell{c}{Learning\\rate} &\tabincell{c}{Batch\\size} & Epoch & Milestones & Gamma & BPP\\
        \hline
        \multirow{7}{*}{JPEG 4:2:0} 
            & \multirow{3}{*}{IDF}
                & RGB 4:4:4 &\multirow{2}{*}{$64\times 64$} & \multirow{3}{*}{1e-4} & \multirow{2}{*}{64} & \multirow{2}{*}{3000} & \multirow{2}{*}{$250,500,750$} & \multirow{3}{*}{0.1} & 6.964 \\
                \cline{3-3} \cline{10-10}
                ~ &~ &YCbCr 4:4:4 &~ &~ &~ &~ &~ &~ & 6.183 \\
                \cline{3-3} \cline{4-4} \cline{6-8} \cline{10-10}
                ~ &~ &DCT 4:2:0 & $256\times 256$ &~ & 32 &1000 & $150$ &~ &1.994 \\ %TODO
            \cline{2-10}
            ~ &\multirow{3}{*}{Multi-scale} 
                & RGB 4:4:4 &\multirow{3}{*}{$256\times 256$} & \multirow{2}{*}{2e-4} & \multirow{3}{*}{64} & \multirow{3}{*}{6000} & \multirow{3}{*}{$3000$} & \multirow{3}{*}{0.5} & 4.398 \\
                \cline{3-3} \cline{10-10}
                ~ &~ &YCbCr 4:4:4 &~ &~ &~ &~ &~ &~ & 3.984 \\
                \cline{3-3} \cline{5-5} \cline{10-10}
                ~ &~ &DCT 4:2:0 &~ &2e-5 &~ &~ &~ &~ &1.674 \\
            \cline{2-10}
            ~ &Ours &DCT 4:2:0 &$256\times256$ &1e-4 &16 &2000 &$1500$ &0.1 &\textbf{0.965}\\
        \hline
        
        \multirow{7}{*}{JPEG 4:4:4} 
            & \multirow{3}{*}{IDF}
                & RGB 4:4:4 &\multirow{3}{*}{$64\times 64$} & \multirow{3}{*}{1e-4} & \multirow{3}{*}{64} & \multirow{3}{*}{3000} & \multirow{3}{*}{$250,500,750$} & \multirow{3}{*}{0.1} & 7.059 \\
                \cline{3-3} \cline{10-10}
                ~ &~ &YCbCr 4:4:4 &~ &~ &~ &~ &~ &~ & 6.362 \\
                \cline{3-3} \cline{10-10}
                ~ &~ &DCT 4:4:4 &~ &~ &~ &~ &~ &~ & 5.875 \\
            \cline{2-10}
            ~ &\multirow{3}{*}{Multi-scale} 
                & RGB 4:4:4 &\multirow{3}{*}{$256\times 256$} & \multirow{2}{*}{2e-4} & \multirow{3}{*}{64} & \multirow{2}{*}{6000} & \multirow{3}{*}{$3000$} & \multirow{3}{*}{0.5} & 4.604 \\
                \cline{3-3} \cline{10-10}
                ~ &~ &YCbCr 4:4:4 &~ &~ &~ &~ &~ &~ &4.079 \\
                \cline{3-3} \cline{5-5} \cline{7-7} \cline{10-10}
                ~ &~ &DCT 4:4:4 &~ &2e-5 &~ &3000 &~ &~ &2.600\\ %TODO
            \cline{2-10}
            ~ &Ours &DCT 4:4:4 &$256\times256$ &1e-4 &16 &2000 &$1500$ &0.1 &\textbf{1.122}\\
        \hline
        
    \end{tabular}
    \caption{Experiment settings and results on Kodak dataset for multi-scale and IDF models. Milestones and gamma are parameters of MultiStepLR in PyTorch. All models are trained and evaluated with QP 75.
    }
    \label{tab:lossless}
\end{table*}

\section{More architecture details}
\label{sec:architectures}

\subsection{Hyper-network}
\label{sec:hyper-network}
\begin{figure}[htb]
  \centering
  \begin{subfigure}{0.45\linewidth}
      \centering
      \includegraphics[height = 3cm]{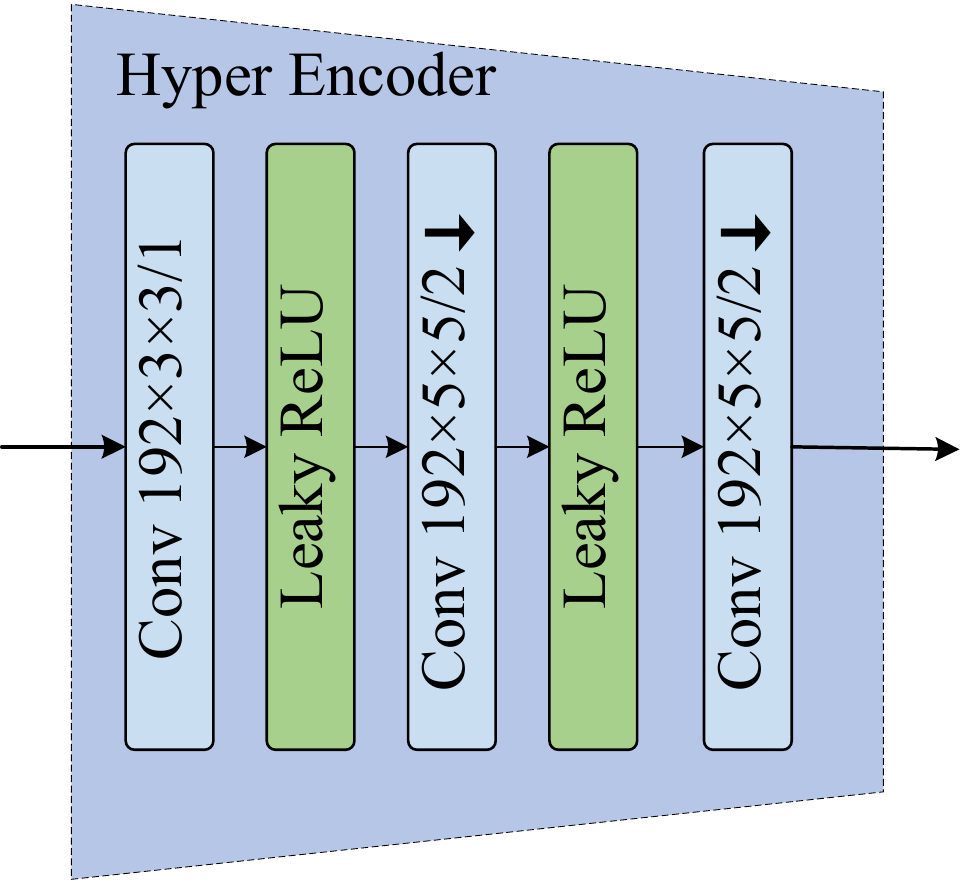}
      \subcaption{Hyper Encoder}
      \label{fig:hyper-encoder}
  \end{subfigure}
  \begin{subfigure}{0.45\linewidth}
      \centering
      \includegraphics[height = 3cm]{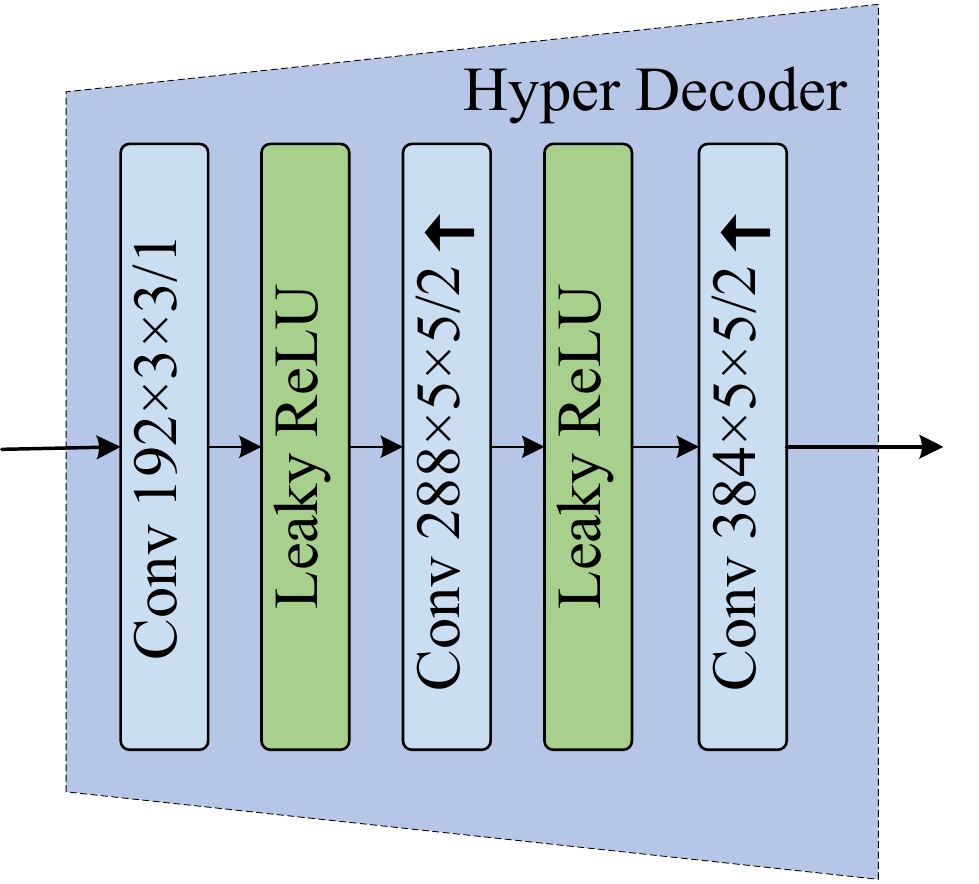}
      \subcaption{Hyper Decoder}
      \label{fig:hyper-decoder}
  \end{subfigure}
  \caption{Detailed architecture of Hyper Encoder and Hyper Decoder.}
  \label{fig:hyper}
\end{figure}

\cref{fig:hyper} shows the architecture details for the hyper-network mentioned in the main text.

\subsection{Cross-color Cases}
\label{ssec:cross-color}
\begin{figure*}[htb]
  \centering
  \includegraphics[width=0.98\linewidth]{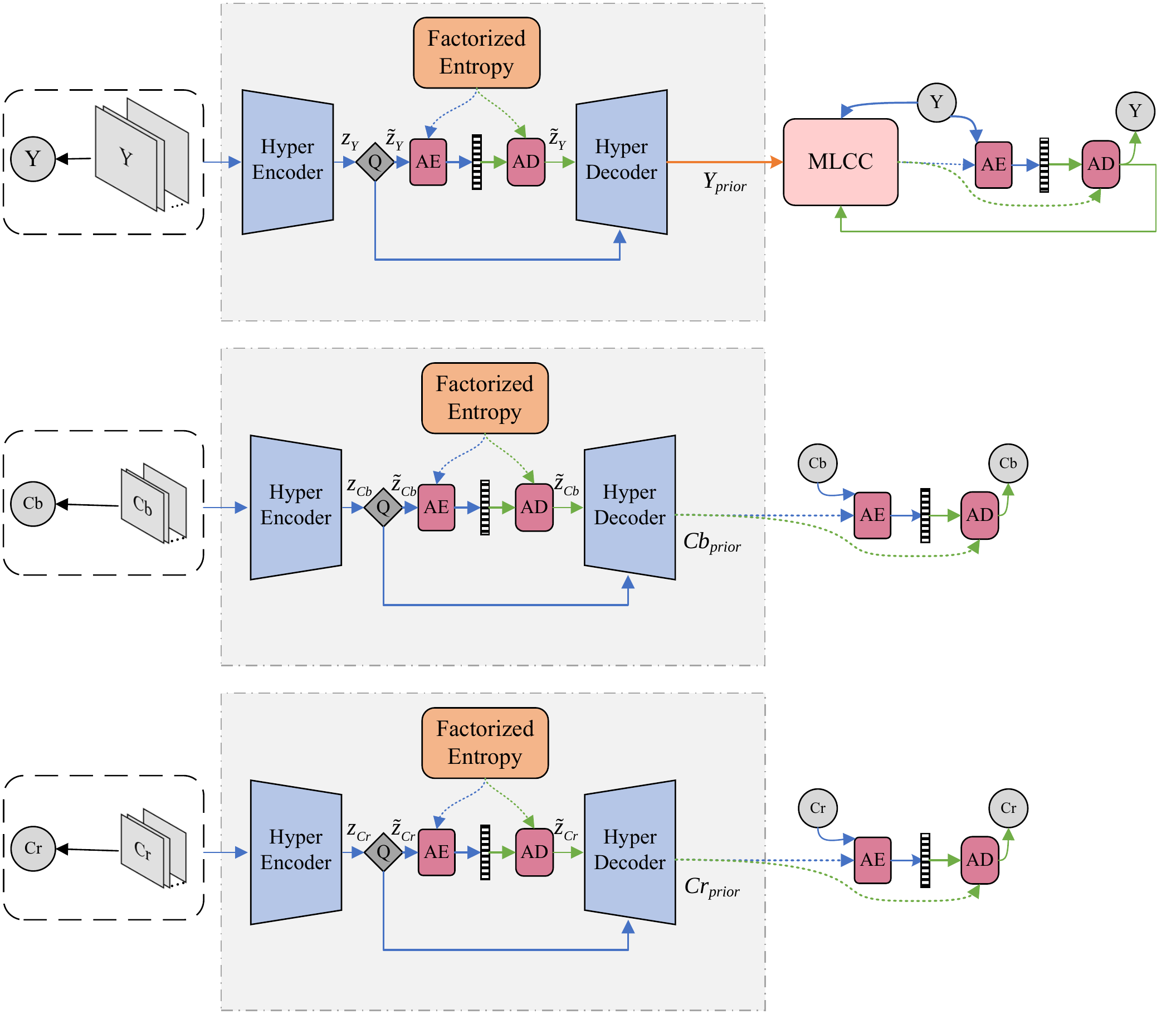}
  \caption{Architecture details of \textbf{Cross-color case 1}. Blue and green lines indicate data-flow for encoding and decoding respectively, orange lines are shared.}
  \label{fig:case1}
\end{figure*}

\begin{figure*}[htb]
  \centering
  \includegraphics[width=0.98\linewidth]{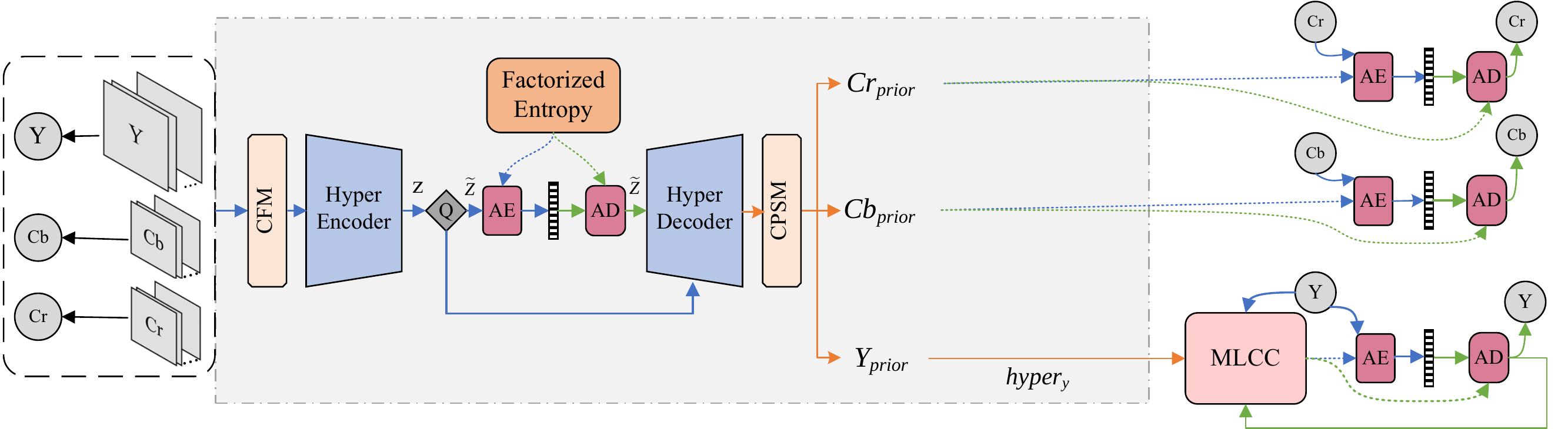}
  \caption{Architecture details of \textbf{Cross-color case 2}. Blue and green lines indicate data-flow for encoding and decoding respectively, orange lines are shared.}
  \label{fig:case2}
\end{figure*}

\begin{figure*}[htb]
  \centering
  \includegraphics[width=0.98\linewidth]{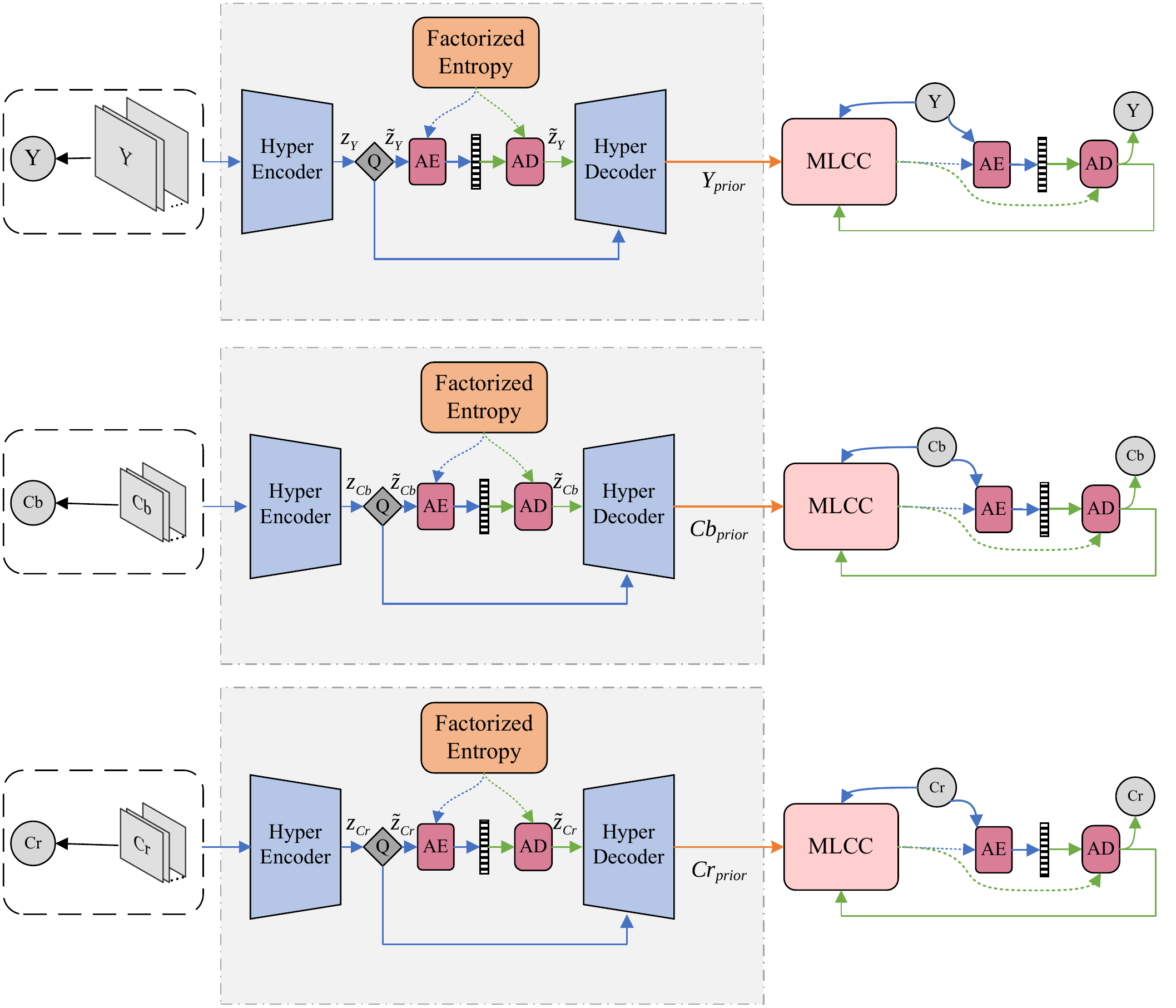}
  \caption{Architecture details of \textbf{Cross-color case 3}. Blue and green lines indicate data-flow for encoding and decoding respectively, orange lines are shared.}
  \label{fig:case3}
\end{figure*}

\cref{fig:case1}, \cref{fig:case2} and \cref{fig:case3} show the architectures for the three cases of cross-color entropy model respectively.

\subsection{Variants of MLCC model}
\label{sec:variants}
\begin{figure*}[htb]
  \centering
  \includegraphics[width=0.98\linewidth]{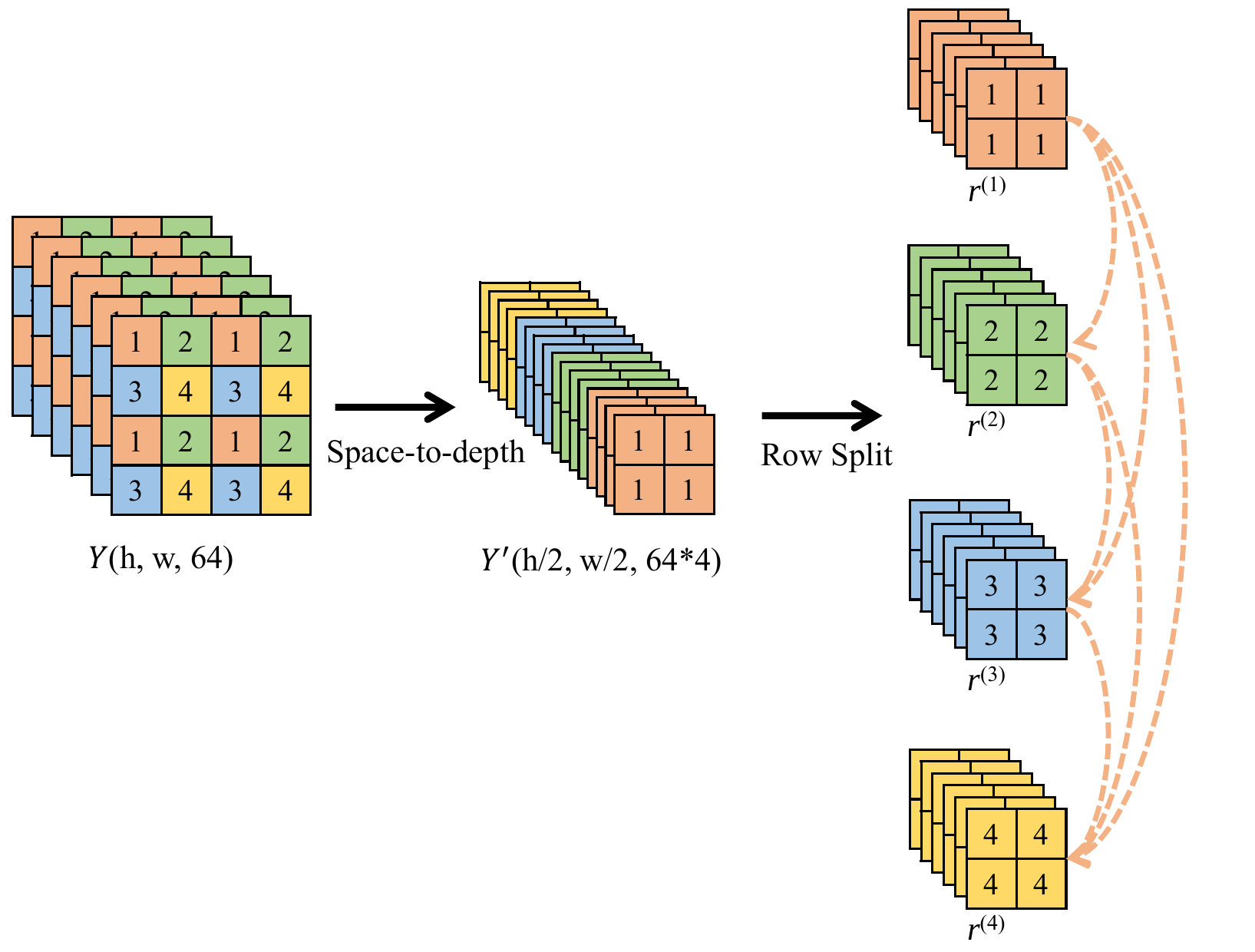}
  \caption{Matrix context modeling method with only row split. Solid arrows indicate data operation and dotted arrows denote conditional relationships.}
  \label{fig:only-row-split}
\end{figure*}

\begin{figure*}[htb]
  \centering
  \includegraphics[width=0.98\linewidth]{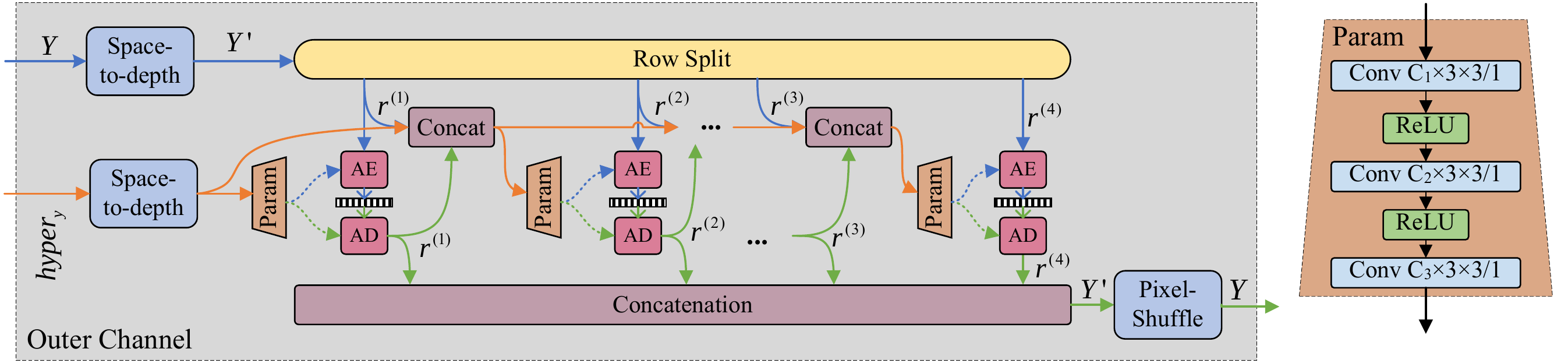}
  \caption{Architecture details of \textbf{Only Outer Channel}. \cref{fig:only-row-split} is its corresponding context model. Letting $n$, $m$ representing the channel number of input tensor and next slice to be modeled respectively, $C_1$, $C_2$ and $C_3$ are decided by $ C_1 = n - d, C_2 = n - 2*d, C_3=2*m, d = (n -2*m)//3$. Blue and green lines indicate data-flow for encoding and decoding respectively, orange lines are shared. }
  \label{fig:only-row}
\end{figure*}

\begin{figure*}[htb]
  \centering
  \includegraphics[width=0.98\linewidth]{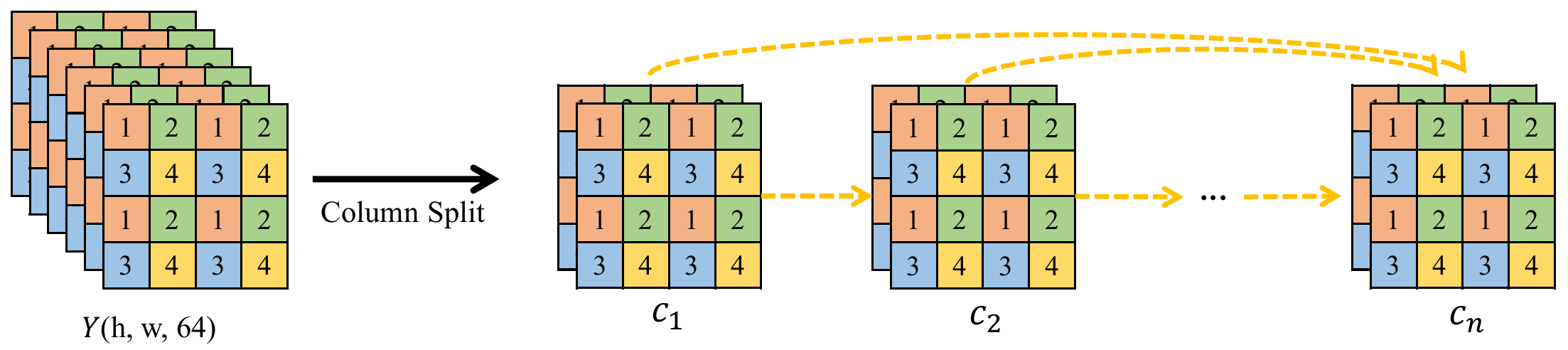}
  \caption{Matrix context modeling method with only column split.Solid arrows indicate data operation and dotted arrows denote conditional relationships.}
  \label{fig:only-column-split}
\end{figure*}

\begin{figure*}[htb]
  \centering
  \includegraphics[width=0.98\linewidth]{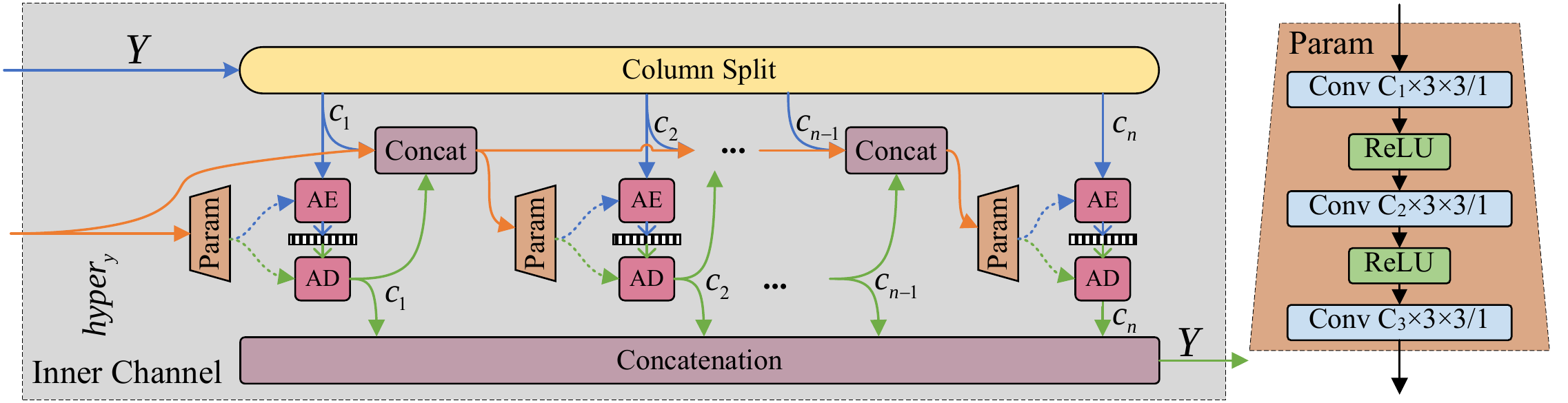}
  \caption{Architecture details of \textbf{Only Inner Channel}. \cref{fig:only-column-split} is its corresponding context model. Letting $n$, $m$ representing the channel number of input tensor and next slice to be modeled respectively, $C_1$, $C_2$ and $C_3$ are decided by $ C_1 = n - d, C_2 = n - 2*d, C_3=2*m, d = (n -2*m)//3$. Blue and green lines indicate data-flow for encoding and decoding respectively, orange lines are shared.}
  \label{fig:only-column}
\end{figure*}

\begin{figure*}[htb]
  \centering
  \includegraphics[width=0.98\linewidth]{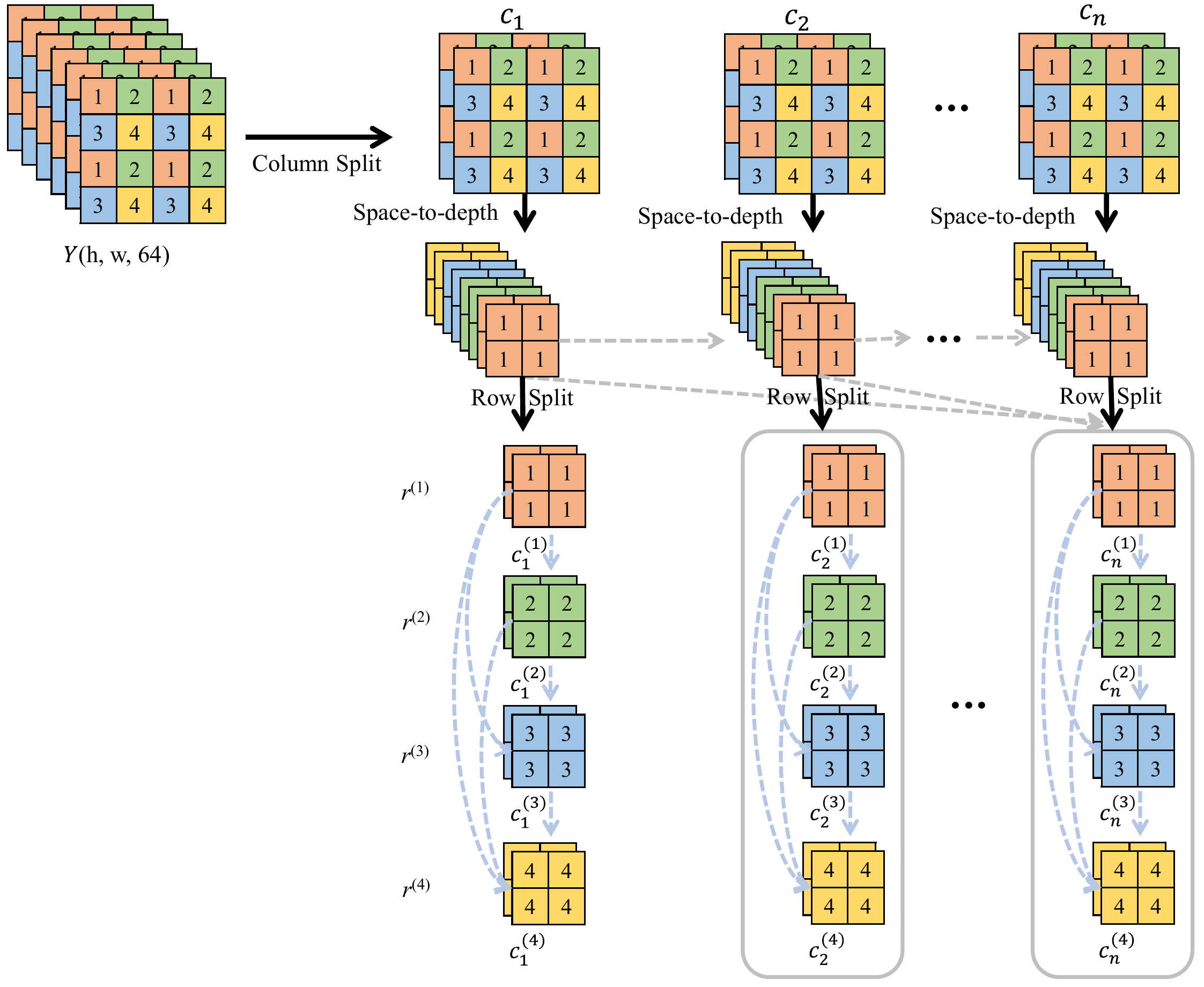}
  \caption{Matrix context modeling method in column-to-row manner.Solid arrows indicate data operation and dotted arrows denote conditional relationships. Light grey and light blue dotted arrows align with Outer Channel and Inner Channel in \cref{fig:column-to-row-mlcc}, respectively.}
  \label{fig:column-to-row}
\end{figure*}

\begin{figure*}[htb]
  \centering
  \includegraphics[width=0.98\linewidth]{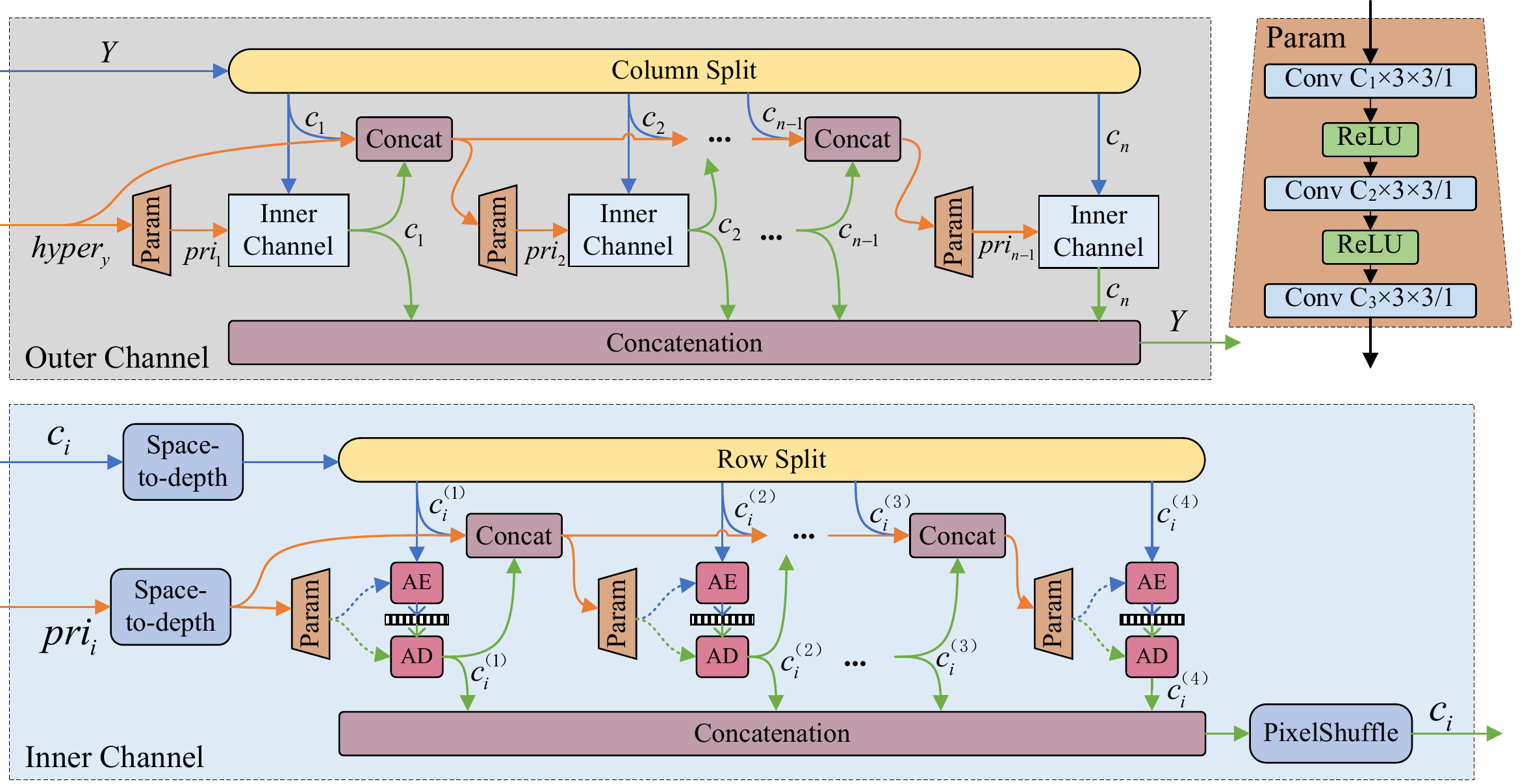}
  \caption{Architecture details of \textbf{column-to-row}. \cref{fig:column-to-row} is its corresponding context model. Letting $n$, $m$ representing the channel number of input tensor and next slice to be modeled respectively, $C_1$, $C_2$ and $C_3$ are decided by $ C_1 = n - d, C_2 = n - 2*d, C_3=2*m, d = (n -2*m)//3$.
  Blue and green lines indicate data-flow for encoding and decoding respectively, orange lines are shared.}
  \label{fig:column-to-row-mlcc}
\end{figure*}

This section shows architecture details of the three variants of MLCC (\ie \textbf{Only Outer Channel} in \cref{fig:only-row}, \textbf{Only Inner Channel} in \cref{fig:only-column}, and \textbf{Column-to-row} in  \cref{fig:column-to-row-mlcc}) and their corresponding context models (\cref{fig:only-row-split}, \cref{fig:only-column-split}, \cref{fig:column-to-row}) mentioned in Sec.\ 4.3 in the main text. %Actually, both row split and column split in our method are used to generate channel slices for channel conditional modelling (\ie \emph{Outer Channel} or \emph{Inner Channel}). %Since row split corresponds to channel dimension, space-to-depth operation, converting the spatial dimension into the channel dimension, is required to add to row split.

%%%%%%%%% REFERENCES
{\small
\bibliographystyle{ieee_fullname}
\bibliography{sup}
}